\documentstyle[preprint,tighten,eqsecnum,aps,floats,psfig,epsfig]{revtex}

\begin{document}
\draft
\title{
Crossover behavior in three-dimensional dilute spin systems 
}
\author{Pasquale Calabrese,$^1$ Pietro Parruccini,$^2$ Andrea Pelissetto,$^3$ 
Ettore Vicari$^2$ }
\address{$^1$ Scuola Normale Superiore and  INFN, Piazza dei Cavalieri 7,
 I-56126 Pisa, Italy.}
\address{$^2$
Dip. Fisica dell'Universit\`a di Pisa
and INFN, V. Buonarroti 2, I-56127 Pisa, Italy}
\address{$^3$ Dip. Fisica dell'Universit\`a di Roma ``La Sapienza" \\
and INFN, P.le Moro 2, I-00185 Roma, Italy}
\address{
\bf e-mail: \rm 
{\tt calabres@df.unipi.it,}
{\tt parrucci@df.unipi.it,} 
{\tt Andrea.Pelissetto@roma1.infn.it},
{\tt vicari@df.unipi.it}
}

\date{\today}

\maketitle

\begin{abstract}
We study the crossover behaviors that can be observed 
in the high-temperature phase of three-dimensional dilute spin systems,
using a field-theoretical approach. In particular, for randomly dilute Ising 
systems we consider the Gaussian-to-random and the pure-Ising-to-random 
crossover, determining the corresponding crossover functions
for the magnetic susceptibility and the correlation length.
Moreover, for the physically interesting cases
of dilute Ising, XY, and Heisenberg systems, we estimate several universal
ratios of scaling-correction amplitudes entering the
high-temperature Wegner expansion of the magnetic susceptibility,
of the correlation length, and of the zero-momentum quartic couplings.
\end{abstract}



\section{Introduction.}
\label{intro}

The critical behavior of randomly dilute magnetic materials is 
of considerable theoretical and experimental interest 
\cite{Aharony-76,Stinchcombe-83,Belanger-00,PV-r,FHY-01}.
A simple model describing these systems is provided by 
the Hamiltonian
\begin{equation}
{\cal H}_p = J\,\sum_{<ij>}  \rho_i \,\rho_j \; s_i \cdot s_j,
\label{latticeH}
\end{equation}
where the sum is extended over all nearest-neighbor sites,
$s_i$ are $M$-component spin variables, and
$\rho_i$ are uncorrelated quenched random variables, which are equal to one 
with probability $p$ (the spin concentration) and zero with probability $1-p$
(the impurity concentration). For sufficiently low dilution $1-p$, i.e. 
above the percolation threshold of the spins, 
the system described by the Hamiltonian ${\cal H}_p$ undergoes a second-order 
phase transition at $T_c(p) < T_c(p=1)$.

The nature of the transition is rather well established. 
In the case of the random Ising model (RIM) corresponding to $M=1$,
the transition belongs to a new universality class which is distinct from the 
Ising universality class describing the critical behavior of the 
pure system. This has been clearly observed in experiments 
\cite{Belanger-00} on dilute uniaxial antiferromagnets, such as 
Fe$_x$Zn$_{1-x}$F${}_2$ and  Mn$_x$Zn$_{1-x}$F${}_2$,
in the absence of magnetic field \cite{zerofield}
and in Monte Carlo simulations of the RIM,
see, e.g., Refs.~\cite{CMPV-03,BFMMPR-98,WD-98,Heuer-93}.
The critical exponents are independent of
the impurity concentration and definitely different from those
of the pure Ising universality class.
Field-theoretical (FT) studies 
\cite{PV-00,th2002,PS-00,FHY-00,SAS-97,TMVD-02} confirm these results.
The fixed point (FP) related to the pure Ising universality class
is unstable with respect to the addition of impurities and
the renormalization-group (RG) flow is driven towards a 
new stable random FP that controls the critical behavior.

Unlike Ising systems, multicomponent O($M$)-symmetric spin systems 
do not change their asymptotic critical behavior
in the presence of random impurities. 
Indeed, according to the  Harris criterion \cite{Harris-74},
the addition of impurities to a system which undergoes 
a continuous transition does not change the critical behavior if the 
specific-heat critical exponent $\alpha$ of the pure system is negative,
as is the case for any $M\ge 2$.
From the point of view of RG theory, the Wilson-Fisher FP of the pure 
O($M$) theory is stable under random dilution. 
The presence of impurities affects only the approach
to the critical regime, giving rise to scaling corrections
behaving as $|\tau|^{\Delta_1}$, where 
$\tau$ is the reduced temperature and $\Delta_1=-\alpha$.
The exponent $\Delta_1$ is rather small
for the physically relevant cases $M=2$ and $M=3$---$\alpha=-0.0146(8)$ 
(Ref.~\cite{CHPRV-01}) and $\alpha=-0.1336(15)$
(Ref.~\cite{CHPRV-02}), respectively---giving rise 
to very slowly decaying scaling corrections.
Experiments on ${}^4$He in porous materials \cite{YC-97,ZR-99} and on
randomly dilute isotropic magnetic materials, 
see, e.g., Refs.~\cite{Kaul-85,KR-94,BK-97},
show that the critical exponents of XY and Heisenberg systems 
are unchanged by disorder (see also the list of results reported
in Ref.~\cite{PV-r}).
But, in order to observe the correct exponents in magnetic systems in 
which the reduced temperature is usually not smaller than $10^{-3}$, it is
important to keep into account the scaling corrections 
in the analysis of the experimental data \cite{Kaul-85}.

In this paper we study the crossover behaviors 
that can be observed in the high-temperature phase of 
three-dimensional dilute spin systems. 
First, we consider the crossover from the Gaussian FP to the stable FP of
the model, i.e.~the random FP for $M=1$ and the pure $O(M)$-symmetric
FP for $M\ge 2$.
Such a crossover can be observed at fixed impurity concentration by varying the 
temperature. If $|T-T_c|/T_c > G$, where $G$ is an appropriate Ginzburg number
\cite{Ginzburg-60}, fluctuations are irrelevant and mean-field behavior is 
expected, while for $|T-T_c|/T_c < G$ the asymptotic critical 
behavior sets in. This crossover is not universal. Nonetheless,
there are limiting situations in which the crossover functions 
become independent of the microscopic details of the statistical system:
This is the case of the critical crossover limit of 
systems with medium-range interactions, i.e.~of systems in which the 
interaction scale is larger than the typical microscopic scale 
\cite{crossover}. In this limit the crossover functions
can be computed by using FT methods: for $O(M)$ models 
precise results have been obtained in Ref.~\cite{BB-85} by using the 
three-dimensional massive scheme and in Refs.~\cite{SD-89,KSD-90}
by using the minimal-subtraction scheme without $\epsilon$ expansion.

In Ising systems there is also another interesting crossover 
associated with the RG flow from the pure Ising FP to the random FP. 
When the concentration $p$ is close to 1, by decreasing the temperature
at fixed $p$, one first observes Ising critical behavior, then a crossover 
sets in, ending with the expected random critical behavior. 
In a suitable limit in which  $p\to 1$ this crossover is universal.
The corresponding universal crossover functions
can be computed by using FT methods.

These crossover behaviors are investigated here by using the 
fixed-dimension perturbative approach in powers
of appropriate zero-momentum quartic couplings. 
We determine the RG trajectories
and the crossover functions of the magnetic susceptibility 
$\chi$ and of the second-moment correlation length $\xi$,
defined from the two-point function
\begin{equation}
G(x) \equiv \overline{\langle \rho_0 \, \rho_x \; s_0 \cdot s_x \rangle} ,
\end{equation}
where the overline indicates the average over dilution and 
$\langle \; \; \rangle$ indicates the sample average at fixed disorder.
This study allows us to compute the corresponding effective exponents and to 
determine several universal ratios of scaling-correction amplitudes entering 
their high-temperature Wegner expansions.
Beside $\chi$ and $\xi$, we also consider zero-momentum quartic correlations
and appropriate combinations that have
a universal high-temperature critical limit, such as 
\begin{eqnarray}
G_4 &\equiv&  - {3 M\over M+2} \lim_{\tau\to 0^+}
        {\chi_4 \over \xi^3 \chi^2}, \nonumber \\
G_{22} &\equiv& - \lim_{\tau\to 0^+} {\chi_{22} \over \xi^3 \chi^2}, \label{Gdef} 
\end{eqnarray}
where $\tau$ is the reduced temperature,
$\chi_4$ is  the zero-momentum four-point connected
correlation function averaged over dilution,
i.e., setting $\mu \equiv \sum_x \rho_x s_x$,
\begin{equation}
V \chi_4 = 
\overline{ 
\langle (\mu\cdot\mu)^2 \rangle - \case{M+2}{M} \, 
       \langle \mu\cdot\mu \rangle^2 
}, 
\end{equation}
and $\chi_{22}$ is defined by 
\begin{equation}
V \chi_{22} = 
\overline{\langle \mu\cdot\mu \rangle^2} 
-  \overline{\langle \mu\cdot\mu \rangle}^2 .
\end{equation}
Their high-temperature Wegner expansion is given by
\begin{eqnarray}
&&\chi = \chi_\tau \tau^{\,-\gamma} 
\left( 1 + \chi_{\tau,1} \tau^{\,\Delta_1} + \chi_{\tau,2} \tau^{\,\Delta_2}
+ ...\right),
\label{corrsc}\\
&&\xi = \xi_\tau \tau^{\,-\nu} 
\left( 1 + \xi_{\tau,1} \tau^{\,\Delta_1} +
\xi_{\tau,2} \tau^{\,\Delta_2} + ...\right),
\label{corrxi}\\
&&G_\# = G_\#^* \left( 1 + G_{\#,\tau,1} \tau^{\,\Delta_1} +
G_{\#,\tau,2} \tau^{\,\Delta_2} + ...\right),
\label{GWE}
\end{eqnarray}
where $\Delta_{1,2}$ are the exponents
associated with the first two independent scaling corrections.
For dilute Ising systems, a recent Monte Carlo study \cite{BFMMPR-98} 
provided the estimate $\Delta_1 = 0.25(3)$;
a rough estimate of $\Delta_2$ is 
$\Delta_2 = 0.55(15)$, cf.~Sec.~\ref{resisi}.
For XY and Heisenberg systems $\Delta_1 =-\alpha$,
while $\Delta_2$ coincides with the leading correction-to-scaling
exponent of the pure model, $\Delta_2 = 0.53(1)$ for $M=2$ and 
$\Delta_2 = 0.56(2)$ for $M=3$, cf. Ref.~\cite{PV-r}.
The ratios 
$\xi_{\tau,i}/\chi_{\tau,i}$ and $\chi_{\tau,i}/G_{\#,\tau,i}$
for $i=1,2$ are universal. 
Their determination may be useful for the analysis of experimental or
Monte Carlo data.
In Eqs.~(\ref{corrsc}--\ref{GWE}) we only report
the leading term for each correction-to-scaling exponent, but it should 
be noted that there are also corrections proportional to $\tau^{2\Delta_1}$,
$\tau^{3\Delta_1}$, etc., that may be more relevant---this is the case of 
systems with $M\ge 2$---than those with exponent 
$\tau^{\Delta_2}$. 

The crossover behavior in dilute models was already studied in 
Refs.~\cite{JOS-95,FHY-00} in the Ising-like case and
in Ref.~\cite{DFHI-03} for multicomponent systems.
However, Refs.~\cite{JOS-95,FHY-00,DFHI-03}
studied the crossover and computed the related effective exponents 
with respect to the RG flow parameter, 
while we compute effective exponents with respect to the reduced 
temperature, which have a direct physical interpretation.

The paper is organized as follows. 
In Sec.~\ref{FTframework} we discuss the FT approach. We first introduce the 
effective Landau-Ginzburg-Wilson $\phi^4$ Hamiltonian and 
some general definitions. Then, we generalize the approach of 
Ref.~\cite{BB-85} by showing how to compute 
the crossover functions of the magnetic susceptibility and of the
correlation length in terms of an effective temperature.
These exact expressions allow us to determine 
the temperature dependence of several quantities near the 
critical point and, as a consequence, to compute 
some universal ratios of scaling-correction amplitudes
entering the high-temperature Wegner expansion of $\chi$,
$\xi$, $G_4$, and $G_{22}$ for dilute Ising, XY, and Heisenberg systems.
These results are presented in Sec.~\ref{uniratios}.
Finally, in  Sec.~\ref{secres} we extend the computation to the whole 
crossover regime, determining RG trajectories and effective exponents
for Ising, XY, and Heisenberg systems with random dilution.
In the case of Ising systems, we also discuss the
Ising-to-RIM crossover, give analytic expressions for the 
crossover scaling functions---details are reported in App.~\ref{IsiRim}---and 
explicitly compute the 
crossover function associated with the magnetic susceptibility.
In App.~\ref{identities} we prove some useful identities
among the RG functions introduced in the FT approach.

\section{RG trajectories and crossover functions}
\label{FTframework}

\subsection{Definitions}

\begin{figure*}[tb]
\hspace{-1.0cm}
\vspace{-0.5cm}
\centerline{\psfig{width=7truecm,angle=-90,file=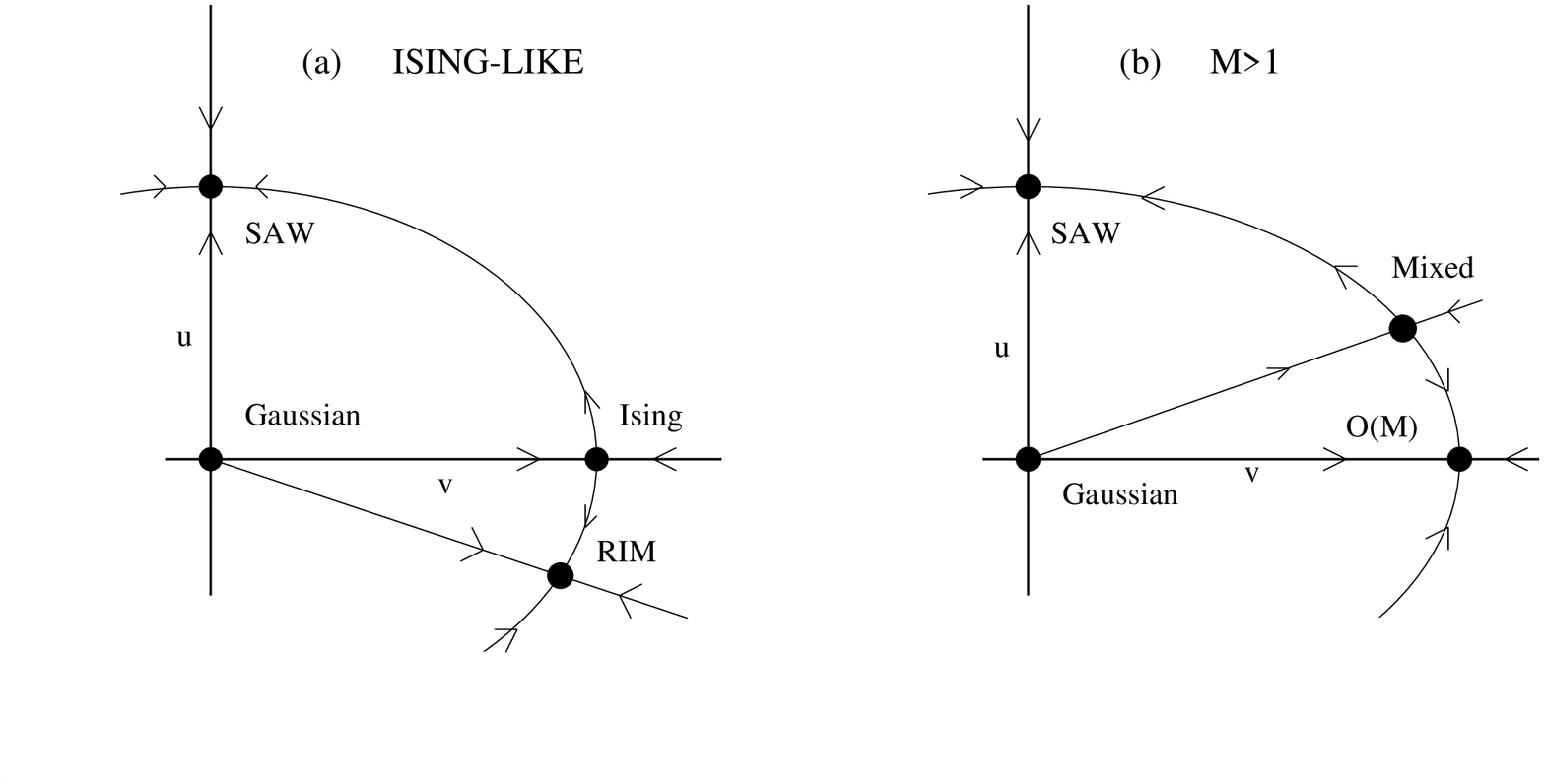}}
\vspace{-0.5cm}
\caption{
Sketch of the RG flow in the coupling plane $(u,v)$ for
(a) Ising ($M=1$) and (b) $M$-component ($M>1$) randomly dilute systems. 
}
\label{randomrgflow}
\end{figure*}

The FT approach is based on an
effective Landau-Ginzburg-Wilson Hamiltonian
that can be obtained by using the replica method 
\cite{Emery-75,EA-75,GL-76,AIM-76}, i.e.
\begin{equation}
{\cal H}_{MN} =  
\int d^d x 
\left\{ \sum_{ia}{1\over 2} \left[ (\partial_\mu \phi_{ai})^2 + 
         r \phi_{ai}^2 \right] + 
  \sum_{ijab} {1\over 4!}\left( u_0 + v_0 \delta_{ij} \right)
          \phi^2_{ai} \phi^2_{bj} 
\right\}, 
\label{Hphi4rim}
\end{equation}
where $a,b=1,...M$ and $i,j=1,...N$.
In the limit $N\rightarrow 0$ the Hamiltonian ${\cal H}_{MN}$ with 
$u_0<0$ and $v_0>0$
is expected to describe the critical properties of dilute
$M$-component spin systems. Thus, their critical behavior can be
investigated by studying the RG flow of ${\cal H}_{MN}$ 
in the limit $N\rightarrow 0$.
For generic values of $M$ and $N$, the Hamiltonian ${\cal H}_{MN}$ describes
$M$ coupled $N$-vector models 
and it is usually called $MN$ model~\cite{Aharony-76}.
${\cal H}_{MN}$ is bounded from below for
$Nu_0 + v_0 > 0$ and $u_0 + v_0 > 0$. 
But, as discussed in Ref.~\cite{MG-82},
in the limit $N\rightarrow 0$ the only stability condition
is $v_0>0$.
Figure~\ref{randomrgflow} sketches the expected
flow diagram in the quartic-coupling plane,
for Ising ($M=1$) and multicomponent ($M\ge 2$) systems
in the limit $N\rightarrow 0$. 
The relevant region for dilute systems corresponds to $u<0$ 
and thus the relevant stable FP is the random FP 
(RIM in Fig.~\ref{randomrgflow}) for $M=1$ and the O($M$) FP for 
$M\ge 2$.

The most precise FT results have been obtained
in the framework of the fixed-dimension expansion  in powers of zero-momentum
quartic couplings. In this scheme the theory is renormalized by introducing
a set of zero-momentum conditions for the one-particle irreducible 
two-point and four-point correlation functions:
\begin{equation}
\Gamma^{(2)}_{ai,bj}(p) =
  \delta_{ai,bj} Z_\phi^{-1} \left[ m^2+p^2+O(p^4)\right],
\label{ren1}  
\end{equation}
where $\delta_{ai,bj} \equiv \delta_{ab}\delta_{ij}$, and
\begin{equation}
\Gamma^{(4)}_{ai,bj,ck,dl}(0) = 
Z_\phi^{-2} m \left( u S_{ai,bj,ck,dl} + v C_{ai,bj,ck,dl} \right),
\label{ren2}  
\end{equation}
where
\begin{eqnarray}
S_{ai,bj,ck,dl} &=& {1\over 3} 
\left(\delta_{ai,bj}\delta_{ck,dl} + \delta_{ai,ck}\delta_{bj,dl} + 
      \delta_{ai,dl}\delta_{bj,ck} \right),
\nonumber \\
C_{ai,bj,ck,dl} &=& {1\over 3}\,\delta_{ij}\delta_{ik}\delta_{il}\,
\left(\delta_{ab}\delta_{cd} + \delta_{ac}\delta_{bd} + 
      \delta_{ad}\delta_{bc} \right).
\end{eqnarray}
In addition one defines the function $Z_t$ through the relation
\begin{equation}
\Gamma^{(1,2)}_{ai,bj}(0) = \delta_{ai,bj} Z_t^{-1},
\label{zt}
\end{equation}
where $\Gamma^{(1,2)}$ is the one-particle irreducible
two-point function with an insertion of $\case{1}{2}\sum_{ai} \phi_{ai}^2$.

The critical behavior is determined by the stable FP of the theory, 
i.e. by the 
common zero $u^*$, $v^*$ of the $\beta$-functions
\begin{equation}
\beta_u(u,v) = \left. m{\partial u\over \partial m}\right|_{u_0,v_0} ,\qquad
\beta_v(u,v) = \left. m{\partial v\over \partial m}\right|_{u_0,v_0} ,
\end{equation}
whose stability matrix has positive eigenvalues
(actually a positive real part is sufficient).
The critical exponents are obtained by evaluating the RG
functions
\begin{equation}
\eta_\phi(u,v) = {\partial \ln Z_\phi\over \partial \ln m},\qquad
\eta_t(u,v) = {\partial \ln Z_t \over  \partial \ln m}
\label{RGfunc}
\end{equation}
at $u^*,v^*$:
\begin{equation}
\eta = \eta_\phi(u^*,v^*),
\qquad
{1\over \nu} = 2 - \eta_\phi(u^*,v^*) + \eta_t(u^*,v^*).
\end{equation}
The six-loop pertubative expansions of the $\beta$ functions and of 
the critical exponents are reported in Refs.~\cite{PV-00,CPV-00}.

In the $MN$ model, the RG functions satisfy a number of identities.
Along the $u=0$ axis we have 
\begin{eqnarray}
&&
\left. {\partial \beta_u\over \partial u} \right|_{u=0}
- \left. {\partial \beta_v\over \partial v} \right|_{u=0} +
\left. {\partial \beta_v\over \partial u} \right|_{u=0}
=0,
\label{identity} \\
&&\left. {\partial \eta_\phi\over \partial u} \right|_{u=0}
-\left. {\partial \eta_\phi\over \partial v} \right|_{u=0}
=0,
\label{etaidentity}
\end{eqnarray}
while along the $v=0$ axis we obtain
\begin{eqnarray}
&&\left. {\partial \beta_u\over \partial u} \right|_{v=0}
- \left. {\partial \beta_v\over \partial v} \right|_{v=0}
- {MN+2\over M+2}\, \left. {\partial \beta_u\over \partial v} \right|_{v=0} =0,
\label{identityv} \\
&&\left. {\partial \eta_\phi\over \partial u} \right|_{v=0}
-{MN+2\over M+2}\,\left. {\partial \eta_\phi\over \partial v} \right|_{v=0}
=0,
\label{etaidentityv}\\
&&\left. {\partial \eta_t\over \partial u} \right|_{v=0}
-{MN+2\over M+2}\,\left. {\partial \eta_t\over \partial v} \right|_{v=0}
=0.
\label{etatidentityv}
\end{eqnarray}
These identities can be proved order by order in the pertubative expansion,
see App.~\ref{identities}.  The second set of relations was already reported 
in Ref.~\cite{CPV-03} for $M=1$.

In the limit $N\rightarrow 0$,
the perturbative expansions in powers of $u$ and $v$ are not
Borel summable at fixed ratio $u/v$ (Ref.~\cite{BMMRY-87} shows it 
explicitly for the zero-dimensional theory with $M=1$, but the
argument has general validity), except when $u=0$
that corresponds to the O($M$)-symmetric  $\phi^4$ theory.
For $M\ge 2$, this is a minor problem since the relevant FP 
is the O($M$)-symmetric one. On the other hand, this is a notable limitation 
of the perturbative approach for the RIM.  Nevertheless, rather reliable 
results for the critical exponents of the RIM universality class have been 
obtained from the analysis of properly resummed perturbative series. 
Several methods have been used:
the Pad\'e-Borel method at fixed $u/v$ or the strictly related 
Chisholm-Borel method, the direct conformal-mapping method,
an expansion around the Ising FP \cite{CMPV-03}, the double-Pad\'e-Borel
and the conformal-Pad\'e-Borel method \cite{PV-00}, which, at least in zero
dimensions \cite{AMR-99}, are able to treat correctly the non-Borel
summability of the expansions at fixed $u/v$. The FT estimates of the 
critical exponents obtained from the analysis of the six-loop expansions
reported in Refs.~\cite{CPV-00,PV-00} depend only slightly
on the resummation method. 
For instance, Ref.~\cite{PV-00} reports $\nu = 0.673(8)$ and 
$\eta = 0.029(3)$ from the direct conformal-mapping method, and 
$\nu = 0.678(10)$ and $\eta = 0.030(3)$ from an analysis that follows the 
ideas of Ref.~\cite{AMR-99}. A second source of 
uncertainty is the position of the FP. Monte Carlo \cite{CMPV-03} 
simulations give $u^* = -18.6(3)$ and $v^* = 43.3(2)$, which are 
significantly different from 
the FT estimates \cite{PV-00} $u^* = -13(2)$ and $v^* = 38.0(1.5)$,
obtained from the numerical determination of the stable common
zero of the $\beta$-functions.
However, as discussed in Ref.~\cite{CMPV-03}, the critical-exponent estimates
show a relatively small dependence on the position of the FP.  By using the 
Monte Carlo results for the location of the FP in the $u$-$v$ plane,
one obtains \cite{CMPV-03} $\nu = 0.686(4)$ and 
$\eta = 0.026(3)$, which are close to the above-reported ones,
obtained by using the field-theoretical estimates of the  FP. 
In any case, it is reassuring that the FT results
are in satisfactory agreement with the Monte Carlo estimates 
of the critical exponents,
i.e. \cite{CMPV-03} $\nu=0.683(3)$ and $\eta=0.035(2)$.
The comparison of the different analyses shows that all different resummation 
methods give results of similar accuracy. In particular, the more 
sophisticated analyses suggested in Ref. \cite{AMR-99} and employed in 
Ref.~\cite{PV-00} apparently do not provide more accurate results than those 
at fixed $u/v$. For this reason, in the following we only use the 
Pad\'e-Borel and the conformal-mapping method at fixed $u/v$.
In the latter case, for the singularity of the Borel transform we use
the naive analytic continuation for $N\to 0$ of the result for the 
cubic model reported in Ref.~\cite{CPV-00}. The results of 
Ref.~\cite{BMMRY-87} suggest that this should allow us to take into 
account the leading divergent behavior of the series 
at least for sufficiently small $|u/v|$ 
(in zero dimensions for $-1/2 < u/v < 0$). 

\subsection{Renormalization-group trajectories}

The RG trajectories in the plane $(u,v)$ are lines which start from the 
Gaussian FP located at $u=v=0$ and along which the quartic Hamiltonian 
parameters $u_0$ and $v_0$ are kept fixed.
They are implicitly characterized by the equation
\begin{equation}
F(u,v)\equiv {u Z_u(u,v)\over v Z_v(u,v)} = {u_0\over v_0} \equiv  s.
\end{equation}
RG trajectories can also be determined by solving the differential equations
\begin{eqnarray}
&&-\lambda {d u\over d\lambda} = \beta_u(u(\lambda),v(\lambda)),\nonumber\\
&&-\lambda {d v\over d\lambda} = \beta_v(u(\lambda),v(\lambda)),
\label{rgflow}
\end{eqnarray}
where $\lambda\in [0,\infty)$, with the initial conditions
\begin{eqnarray}
&&u(0) = v(0) = 0 ,\nonumber \\
&& \left. {d u\over d\lambda} \right|_{\lambda=0} = s ,\qquad
\left. {d v\over d\lambda} \right|_{\lambda=0} = 1. \label{ini-rgflow}
\end{eqnarray}
The solutions $u(\lambda,s)$ and $v(\lambda,s)$ provide the RG trajectories 
in the $(u,v)$ plane as a function of $s$.
The RG trajectories relevant for dilute spin systems are those with $s<0$.
The attraction  domain of the stable FP is given by 
the values of $u_0$ and $v_0$ corresponding to trajectories 
ending at the stable FP, i.e. trajectories for which 
\begin{equation}
u(\lambda=\infty,s) = u^*, \qquad v(\lambda=\infty,s) = v^*.
\end{equation}
The crossover functions from the Gaussian to the Wilson-Fisher
stable FP have been much studied in the case of the O($M$)-symmetric theories,
both in the field theory \cite{BB-85,SD-89,KSD-90} and 
in medium-range models \cite{crossover}.
In order to determine the crossover functions along the 
RG trajectories, and in particular those related with the correlation 
length $\xi$, the magnetic susceptibility $\chi$, and the reduced
temperature $\tau\propto r-r_c$,
we extend the method of Ref.~\cite{BB-85} to Hamiltonians with many
quartic parameters, such as ${\cal H}_{MN}$. Using the relations
\begin{eqnarray}
&&\xi = 1/m ,\qquad \overline{\chi}\equiv {1\over M}\chi  = Z_\phi \xi^2,
\nonumber \\
&& \Gamma^{(1,2)}_{ai,bj}(0) =
\left. {\partial \Gamma^{(2)}_{ai,bj}(0)\over \partial \tau }\right|_{u_0,v_0} = 
\delta_{ai,bj} \left. {\partial \overline{\chi}^{\,-1}\over \partial \tau }\right|_{u_0,v_0} = 
\delta_{ai,bj} Z_t^{-1},
\end{eqnarray}
and Eq.~(\ref{RGfunc}), and defining
\begin{eqnarray}
&&\hat\eta_\phi(\lambda,s) \equiv \eta_\phi(u(\lambda,s),v(\lambda,s)),
\nonumber \\
&&\hat\eta_t(\lambda,s) \equiv  
\eta_t(u(\lambda,s),v(\lambda,s)) - \eta_\phi(u(\lambda,s),v(\lambda,s)),
\end{eqnarray}
 we derive the following expressions
\begin{eqnarray}
&&\widetilde\xi (\lambda,s)\equiv \xi v_0 = \lambda, \label{flxi} \\ 
&& \widetilde\chi (\lambda,s)\equiv \overline{\chi} v_0^2 = \lambda^2 
\exp \left[ - \int_0^\lambda d x \,{\hat\eta_\phi(x,s)\over x} \right] ,
\label{flchi} \\ 
&& \widetilde\tau (\lambda,s) = \tau/v_0^2 =
\int_\lambda^\infty d x \, {2 - \hat\eta_\phi(x,s)\over x^3}\,
\exp  \int_0^x dy {\hat\eta_t(y,s)\over y} ,
\label{fltau} 
\end{eqnarray}
where $\widetilde\xi$, $\widetilde\chi$, and $\widetilde\tau$ are dimensionless
quantities.
One can easily verify that in the Gaussian limit, i.e. 
for $\lambda\rightarrow 0^+$ or $\widetilde\tau\rightarrow \infty$, we have
$u,v=O(\lambda)$, 
$\hat\eta_\phi(\lambda,s)=O(\lambda^2)$,
$\hat\eta_t(\lambda,s)=O(\lambda)$;
therefore $\widetilde\tau \widetilde\xi^{\,2}\rightarrow 1$ and
$\widetilde\chi \widetilde\xi^{\,-2}\rightarrow 1$, as expected.

Eqs.~(\ref{flxi}--\ref{fltau}) allow us to compute 
$\widetilde\xi$ and $\widetilde\chi$ as  functions of $\widetilde\tau$ 
and $s$. We can then define effective exponents
by taking logarithmic derivatives of $\widetilde\xi$  and $\widetilde\chi$ at 
fixed $s$:
\begin{equation}
\nu_{\rm eff}(\widetilde{\tau},s) \equiv - 
\left. {\partial {\rm ln} \widetilde\xi \over \partial {\rm ln} \widetilde\tau}
\right|_{s},
\qquad
\gamma_{\rm eff}(\widetilde{\tau},s) \equiv - 
\left. 
{\partial {\rm ln} \widetilde\chi \over \partial {\rm ln} \widetilde\tau}
\right|_{s},
\qquad
\eta_{\rm eff}(\widetilde{\tau},s) \equiv 2 - \left.
{\partial {\rm ln} \widetilde\chi\over \partial {\rm ln} \widetilde\xi}
\right|_s .
\label{effexp}
\end{equation}
One can easily check that $\eta_{\rm eff} = 
2 - \gamma_{\rm eff}/\nu_{\rm eff} = \hat{\eta}_\phi$.
On the other hand, $\gamma_{\rm eff}\not=\gamma(u,v)$ and 
$\nu_{\rm eff}\not=\nu(u,v)$ where $\gamma(u,v)$ and $\nu(u,v)$
are the RG functions associated with the exponents $\gamma$ and $\nu$.

\section{Universal ratios of scaling-correction amplitudes}
\label{uniratios}

\subsection{General results}
\label{genres}

In order to determine the scaling-correction amplitudes, we compute the 
crossover functions close to the critical point, i.e., for 
$\lambda\rightarrow\infty$ or $\widetilde\tau\rightarrow 0^+$.
For this purpose,
we consider the expansion of the RG functions
around the stable FP $(u^*,v^*)$. 
We write
\begin{eqnarray}
&&\beta_u(u,v) \approx  b_{uu} (u-u^*) + b_{uv} (v-v^*),
\nonumber \\
&&\beta_v(u,v) \approx  b_{vu} (u-u^*) + b_{vv} (v-v^*).
\end{eqnarray}
Then, using Eq.~(\ref{rgflow}) we have the following behavior,
in the limit 
$\lambda\rightarrow\infty$
and for values of $s$ in the attraction domain of the stable FP,
\begin{eqnarray}
u(\lambda,s)\approx u^* + u_{\lambda,1}(s) \lambda^{-\omega_1} + 
u_{\lambda,2}(s) \lambda^{-\omega_2}+ ...,
\nonumber \\
v(\lambda,s)\approx v^* + v_{\lambda,1}(s) \lambda^{-\omega_1} + 
v_{\lambda,2}(s) \lambda^{-\omega_2}+ ...,
\label{uvexp}
\end{eqnarray}
where $\omega_1,\omega_2$ are the eigenvalues of the matrix
\begin{equation}
\left(\matrix{
b_{uu} & b_{uv} \cr
b_{vu} & b_{vv} 
}\right), 
\label{omegama}
\end{equation}
and we are keeping only the leading terms in
powers of $\lambda^{-\omega_1}$ and $\lambda^{-\omega_2}$.
In particular, we have neglected terms of order $\lambda^{-2\omega_1}$,
$\lambda^{-3\omega_1}$, etc.,
which may be as important as those of order $\lambda^{-\omega_2}$.
Moreover, we have 
\begin{eqnarray}
R_1 \equiv {u_{\lambda,1}(s)\over v_{\lambda,1}(s) } = 
                     {\omega_1 - b_{vv}\over b_{vu}}
                   = {b_{uv}\over \omega_1- b_{uu}}, \nonumber \\
R_2 \equiv {u_{\lambda,2}(s)\over v_{\lambda,2}(s) } = 
                     {\omega_2 - b_{vv}\over b_{vu}}
                   = {b_{uv}\over \omega_2- b_{uu}}. 
\label{ridef} 
\end{eqnarray}
These ratios are independent of $s$, as expected because 
they are universal.  Indeed, as we shall see, they 
 can be related to the universal ratios $G_{22,\tau,i}/G_{4,\tau,i}$
of the scaling-correction amplitudes of $G_4$ and $G_{22}$,
cf. Eqs.~(\ref{Gdef}) and (\ref{GWE}).

We also expand the RG functions associated with the critical exponents,
\begin{eqnarray}
&&\nu(u,v) \equiv {1\over 2 + \eta_t(u,v) - \eta_\phi(u,v)}
\approx  \nu + \nu_u (u-u^*) + \nu_v (v-v^*), \nonumber \\
&&\gamma(u,v) \equiv \left[ 2 - \eta_\phi(u,v)\right]\nu(u,v) 
\approx  \gamma + \gamma_u (u-u^*) + \gamma_v (v-v^*). 
\end{eqnarray}
and define the $s$-independent quantities
\begin{equation}
\gamma_{\lambda,i} \equiv  \gamma_u R_i + \gamma_v ,\qquad
\nu_{\lambda,i} \equiv \nu_u R_i + \nu_v, \qquad \Delta_i=\omega_i\nu,
\end{equation}
for $i=1,2$.  Then, using Eq.~(\ref{uvexp}), we find
\begin{eqnarray}
&&\widetilde\chi(\lambda,s) = \chi_{\lambda}(s) \lambda^{2-\eta} 
\left( 1 + \sum_{i=1}^2 \chi_{\lambda,i}(s) \lambda^{-\omega_i} +
...\right), 
\nonumber \\
&& \chi_{\lambda}(s) = 
   \exp \left( 
        -\int_0^1 d x \; {\hat\eta_\phi(x,s)\over x} 
        -\int_1^\infty d x \; {\hat\eta_\phi(x,s) - \eta \over x} 
       \right),
\nonumber \\
&&\chi_{\lambda,i}(s) = {\gamma\over \Delta_i} 
\left( {\nu_{\lambda,i}\over \nu} - {\gamma_{\lambda,i}\over \gamma}\right)
v_{\lambda,i}(s),
\label{chila}
\end{eqnarray}
and
\begin{eqnarray}
&&\widetilde\tau(\lambda,s) = \tau_{\lambda}(s) \lambda^{-1/\nu} 
\left( 1 + \sum_{i=1}^2 \tau_{\lambda,i}(s) \lambda^{-\omega_i} +
...\right), 
\nonumber \\
&& \tau_{\lambda}(s) = \gamma \exp \left(
      -\int_0^1 d x \; {\hat\eta_t(x,s)\over x}
      -\int_1^\infty d x \; {\hat\eta_t(x,s)+2-1/\nu\over x}
                                  \right),\nonumber \\
&&\tau_{\lambda,i}(s) = 
\left( {\gamma_{\lambda,i}\over (1+\Delta_i)\gamma} - {\nu_{\lambda,i}\over \Delta_i \nu}
\right) v_{\lambda,i}(s).
\label{taula} 
\end{eqnarray}
Using  Eqs.~(\ref{chila}) and (\ref{taula}), 
we can derive the Wegner expansion of 
$\xi$, $\chi$, and of the zero-momentum quartic couplings $u$ and $v$
in terms of the reduced temperature $\widetilde\tau$. We obtain
\begin{eqnarray}
&&\widetilde\xi(\widetilde\tau,s) = \xi_\tau(s) \widetilde\tau^{\,-\nu} 
\left( 1 + \sum_{i=1}^2 \xi_{\tau,i}(s) \widetilde\tau^{\,\Delta_i} + 
      ...\right),
\nonumber \\
&& \xi_\tau(s) = \tau_\lambda(s)^{\nu}, \qquad 
\xi_{\tau,i}(s) = \nu \tau_{\lambda,i}(s) \,v_{\lambda,i}(s)\, 
\tau_\lambda(s)^{-\Delta_i}, 
\end{eqnarray}
and 
\begin{eqnarray}
&&\widetilde\chi(\widetilde\tau,s) = \chi_\tau(s) \widetilde\tau^{\,-\gamma} 
\left( 1 + \sum_{i=1}^2 \chi_{\tau,i}(s) \widetilde\tau^{\,\Delta_i} + 
      ...\right),
\nonumber \\
&& \chi_\tau(s) = \chi_\lambda(s) \tau_\lambda(s)^{\gamma}, \qquad 
\chi_{\tau,i}(s) = - {\gamma_{\lambda,i}\over \Delta_i ( 1 + \Delta_i)} 
v_{\lambda,i}(s) \tau_\lambda(s)^{-\Delta_i},
\end{eqnarray}
and also
\begin{eqnarray}
&&v(\widetilde\tau,s) = v^* + \sum_{i=1}^2 v_{\tau,i}(s)
\widetilde\tau^{\,\Delta_i} + ...,
\qquad v_{\tau,i}(s) = v_{\lambda,i}(s)\,\tau_\lambda(s)^{-\Delta_i},\\
&&u(\widetilde\tau,s) = u^*  + \sum_{i=1}^2 u_{\tau,i}(s) 
\widetilde\tau^{\,\Delta_i} + ...,
\qquad u_{\tau,i}(s) =  R_i \,v_{\lambda,i}(s)\,\tau_\lambda(s)^{-\Delta_i}.
\end{eqnarray}
The results of Ref.~\cite{CPV-03-eqst} allow us to identify 
\begin{equation}
G_4(\widetilde\tau,s) = v(\widetilde\tau,s),
\qquad 
G_{22}(\widetilde\tau,s) = {1\over 3} u(\widetilde\tau,s),
\end{equation}
and to obtain the corresponding scaling-correction
amplitudes $G_{4,\tau,i}$ and $G_{22,\tau,i}$ defined in Eq.~(\ref{GWE}).

From the above-reported relations we derive the following expressions for 
the universal ratios of scaling-correction amplitudes
\begin{eqnarray}
&&{u_{\tau,i}\over v_{\tau,i}} =  R_i, \nonumber \\
&&{\xi_{\tau,i}\over \chi_{\tau,i}} = 
{\nu_{\lambda,i}(1+\Delta_i)\over\gamma_{\lambda,i}}-
          {\nu\Delta_i\over \gamma},
\nonumber \\
&&{\chi_{\tau,i}\over v_{\tau,i}} =  
-{ \gamma_{\lambda,i}\over \Delta_i (1+\Delta_i) }.
\label{scalcorr-FT}
\end{eqnarray}
Their universality is explicitly verified since they are
independent of $s\equiv u_0/v_0$.

\subsection{Results for dilute Ising systems}
\label{resisi}

Using the results reported in Sec.~\ref{genres}, we can 
estimate several universal scaling-correction amplitude ratios.
We analyze appropriate perturbative series that can be derived from 
those of the $\beta$ functions and the critical exponents.
Again, we use the conformal-mapping method and the Pad\'e-Borel method at 
fixed ratio $u/v$.  The errors we report take into account the 
resummation error and the uncertainty in the location of the FP. We compute
each quantity at the FT and at the Monte Carlo FP. The final error
is such to include both estimates.

As a first step in the analysis we computed the subleading exponents
and the ratios $R_1$ and $R_2$. 
The exponent $\omega_1$ was already computed in Ref.~\cite{PV-00},
obtaining $\omega_1 = 0.25(10)$ (using the double-Pad\'e-Borel and the 
conformal-Pad\'e-Borel method) and $\omega_1 = 0.34(11)$ (using the direct 
conformal-mapping method), in substantial agreement with the Monte Carlo result 
$\omega_1 = 0.37(5)$ of Ref.~\cite{BFMMPR-98}. 
In those analyses the field-theoretical estimates of the FP was used. 
We tried to compute $\omega_1$ 
by also using the Monte Carlo estimate of the FP.
However, all methods gave largely fluctuating results and no estimate 
could be obtained. Then, we determined $\omega_2$.  In this case, 
the conformal-mapping method
provided reasonably stable results up to the Monte Carlo FP. 
We obtained \cite{omega2} $\omega_2 = 0.8(2)$. 

Similar analyses were done for $R_1$ and $R_2$.
Our final results are
\begin{equation}
R_1=-0.90(2),\qquad R_2=-0.7(3).
\end{equation} 
Finally, we determined the ratios of scaling-correction amplitudes
using relations (\ref{scalcorr-FT}). 
In order to have a check of the results, for each quantity we considered
several series with the same FP value. We obtained
\begin{eqnarray}
&&\chi_{\tau,1}/\xi_{\tau,1}  =   1.99(4), \nonumber \\
&&\chi_{\tau,1}/ G_{4,\tau,1} =  -1.0(3), \nonumber \\
&&G_{22,\tau,1}/ G_{4,\tau,1} =   2.1(1),  \nonumber \\
&&\chi_{\tau,2}/\xi_{\tau,2}  =   1.7(2), \nonumber \\
&&\chi_{\tau,2}/ G_{4,\tau,2} =  -0.4(2), \nonumber \\
&&G_{22,\tau,2}/ G_{4,\tau,2} =   1.6(7). 
\end{eqnarray}
The errors take into account the
results obtained from different series and 
different resummation methods, and also the uncertainty on the location of the 
FP. It is interesting to note that the results for the
ratios $\chi_{\tau,i}/\xi_{\tau,i}$ show that 
the quantity $\chi/\xi^2$ has much smaller scaling corrections than 
$\chi$ and $\xi$. This fact was used in Ref.~\cite{CMPV-03} in order to 
obtain a precise Monte Carlo estimate of $\eta$ from
the high-temperature behavior of $\chi/\xi^2$.
For comparison, we report the corresponding values 
for the pure Ising universality class. 
From the analysis of high-temperature series one obtains
$\chi_{\tau,1}/\xi_{\tau,1}=1.11(12)$ (Ref.~\cite{CPRV-02}) and 
$\chi_{\tau,1}/\xi_{\tau,1}=1.32(10)$ (Ref.~\cite{BC-02}), 
while field theory gives \cite{BB-85}
$\chi_{\tau,1}/\xi_{\tau,1}=1.47(4)$ and
$\chi_{\tau,1}/ G_{4,\tau,1} = -0.30(4)$.

\subsection{Results for dilute multicomponent systems}
\label{resmulti}

As in the Ising case, we determine the universal ratios
of scaling-correction amplitudes by analyzing the
six-loop expansions of the $MN$ model \cite{PV-00}.
Since the corresponding RG functions must be evaluated at the
O($M$)-symmetric FP, i.e. along the $u=0$ axis,
the series are Borel summable and 
the standard conformal-mapping method works well.

Identity (\ref{identity}) allows us to obtain the following 
exact results for the universal quantities $R_i$:
\begin{equation}
R_1 = -1 ,\qquad R_2=0,
\end{equation}
which hold independently of $M$.
We also obtain 
\begin{eqnarray}
&&\chi_{\tau,1}/\xi_{\tau,1}  =   1.97(2), \nonumber \\
&&\chi_{\tau,1}/ G_{4,\tau,1} =  -17(2),
\end{eqnarray}
for dilute XY systems, and
\begin{eqnarray}
&&\chi_{\tau,1}/\xi_{\tau,1} =   1.97(2),\nonumber \\
&&\chi_{\tau,1}/ G_{4,\tau,1} =  -2.5(4),
\end{eqnarray}
for dilute Heisenberg systems.
The ratios $\chi_{\tau,2}/\xi_{\tau,2}$ and
$\chi_{\tau,2}/G_{4,\tau,2}$  are just the universal
ratios of scaling-correction amplitudes of the O($M$)-symmetric models.
Ref.~\cite{BB-85} reports
$\chi_{\tau,2}/\xi_{\tau,2}  =   1.57(2)$ and
$\chi_{\tau,2}/\xi_{\tau,2} =   1.63(4)$
respectively for XY and Heisenberg systems.
We add the results
$\chi_{\tau,2}/ G_{4,\tau,2} =  -0.47(5)$ and 
$\chi_{\tau,2}/ G_{4,\tau,2} =   -0.59(5)$ again 
for XY and Heisenberg systems.
We finally mention an $\epsilon$-expansion 
study of the universal ratios of
scaling-correction amplitudes \cite{KG-96}, where the specific-heat 
and low-temperature quantities are considered. These results 
differ significantly from those determined in experiments \cite{KR-94}
on Ni$_{80-x}$Fe$_x$(B,Si)$_{20}$.

\section{Crossovers in randomly dilute spin systems}
\label{secres}

\subsection{Crossover from Gaussian to random critical behavior in Ising 
systems}
\label{crossGRIM}

In the case of the RIM, the FP's have been determined by using 
FT and Monte Carlo methods. For the random FP, we mention again the 
estimates $u^*=-18.6(3)$ and $v^*=43.3(2)$ obtained by Monte Carlo 
simulations \cite{CMPV-03} and the FT results 
reported in Ref.~\cite{PV-00}, $u^*=-13(2)$ and $v^*=38.0(1.5)$.
The position of the unstable Ising FP is $u_I=0$, $v_I=23.56(2)$ 
(Ref.~\cite{CPRV-02}).
The RG trajectories for $s>0$ 
are not interesting for dilute systems; we only mention that
they are attracted by another stable FP with O($N$) symmetry ($N\to 0$),
located at \cite{PV-r,PV-00-g} $u=26.3(4)$, $v=0$. 

\begin{figure}[tb]
\centerline{\psfig{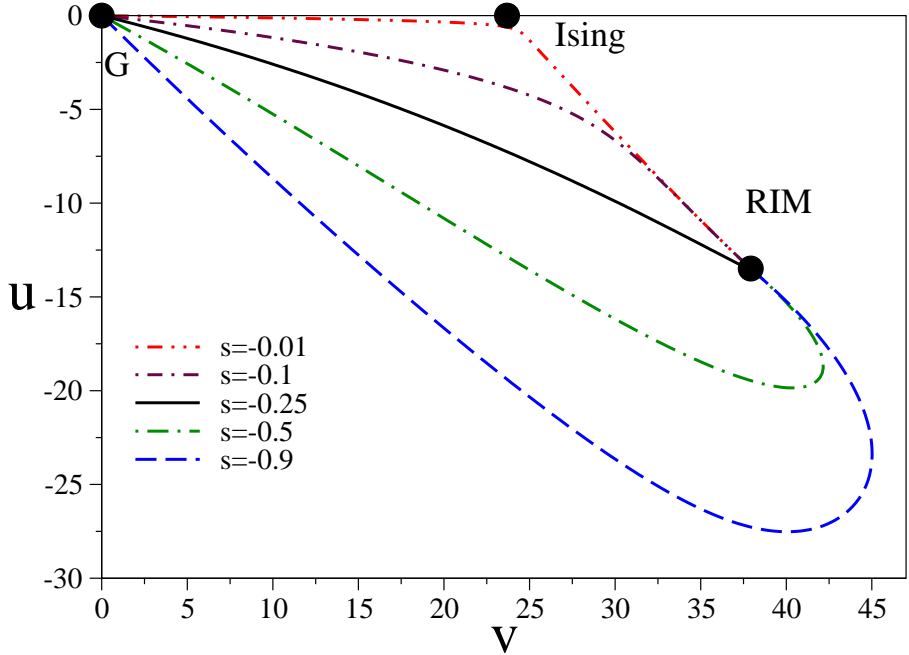}}
\vspace{2mm}
\caption{
Ising systems: 
RG trajectories in the $(u,v)$ plane for several  values of $s$ in the 
interval $-1<s<0$.
}
\label{figtraj}
\end{figure}

In Fig.~\ref{figtraj} we show the RG trajectories
for several values of $s$ in the interval $-1\lesssim s<0$, as obtained by 
numerically integrating the RG equations (\ref{rgflow}),
after resumming the $\beta$-functions.
The figure has been obtained by using a single approximant, but others 
give qualitatively similar results. 
The resummation becomes less and less effective 
as $|s|$ increases. This is expected since the singularities that make the 
perturbative series non-Borel summable play an increasingly important role
as $|s|$ gets larger.
In any case, for $-1 \lesssim s < 0$, the RG 
trajectories flow towards the random FP. 
For $s \lesssim - 1$, Pad\'e-Borel resummations (in this case we cannot 
use the conformal-mapping method since the singularity we use is 
on the positive real axis) hint at runaway RG
trajectories. If this is true and not simply an artifact of the 
perturbative approach, this suggests the existence of a value 
$s_{\rm min}\approx -1$ such that systems corresponding to 
$s<s_{\rm min}$ do not undergo a continuous transition.
As a consequence, since $u$ is directly related to the variance of disorder, 
the continuous transition is expected to disappear for sufficiently large 
disorder.  This prediction may be checked by considering 
a lattice version of the continuum Hamiltonian
\begin{equation}
\int d^3x\; \left[ 
  {1\over2} \left( \partial_\mu\varphi(x)\right) ^2 + {1\over2} 
  \left(t + r(x)\right) \varphi(x)^2 + {g\over 4!} \varphi(x)^4 \right],
\end{equation}
where $\varphi$ is a scalar field and $r(x)$ is a Gaussian uncorrelated random 
variable. Such a model is the starting point of the FT studies of dilute 
systems and, by using the replica trick, can be shown to be equivalent
to the model with Hamiltonian (\ref{Hphi4rim}).
Our results suggest that there is a critical value $v_c$ 
such that, if the variance of $r(x)$ is larger than $v_c$, the continuous 
transition disappears. 

Beside $s=s_{\rm min}$, there is a second interesting value of $s$, 
the value $s^*$
such that the RIM FP is approached from above for $s^* < s < 0$ 
and from below for $s_{\rm min}<s<s^*$. One can easily realize
that for this particular value of $s$ the leading scaling
corrections proportional to $\tau^{\Delta_1}$---and more generally
proportional to $\tau^{n\Delta_1}$---are not present
in the Wegner expansions of the thermodynamic quantities.
Numerically, by using the conformal-mapping method, we 
obtain $s^*= -0.25(5)$.

The behavior of the RG trajectories close 
to the RIM FP can be determined by
using the results presented in Sec.~\ref{genres}. We find that 
$v(\lambda,s)$ can be expanded as
\begin{equation}
v(\lambda,s) \approx v^* + {1\over R_1}(u - u^*) -
      u_{\lambda,2}(s) \left({1\over R_1} - {1\over R_2}\right)
      \left({u-u^*\over u_{\lambda,1}(s)}\right)^{\omega_2/\omega_1},
\label{v-at-RIM}
\end{equation}
where $R_1$ and $R_2$ are universal constants reported in Sec.~\ref{genres},
cf.~Eq.~(\ref{ridef}), and
$u_{\lambda,1}(s)$ and $u_{\lambda,2}(s)$ are expansion coefficients defined
in Eq.~(\ref{uvexp}). Note the presence of the nonanalytic correction
which shows that, close to the FP, trajectories are only defined 
for $(u-u^*)/u_{\lambda,1}(s) > 0$. This is expected on the basis of 
general arguments \cite{Nickel,Sokal-95,PV-nonanal}: 
along any RG trajectory one expects nonanalytic 
corrections proportional, for instance, to $n \omega_i/\omega_1 + m$, 
$n,m$ being integers.

\begin{figure}[tb]
\centerline{\psfig{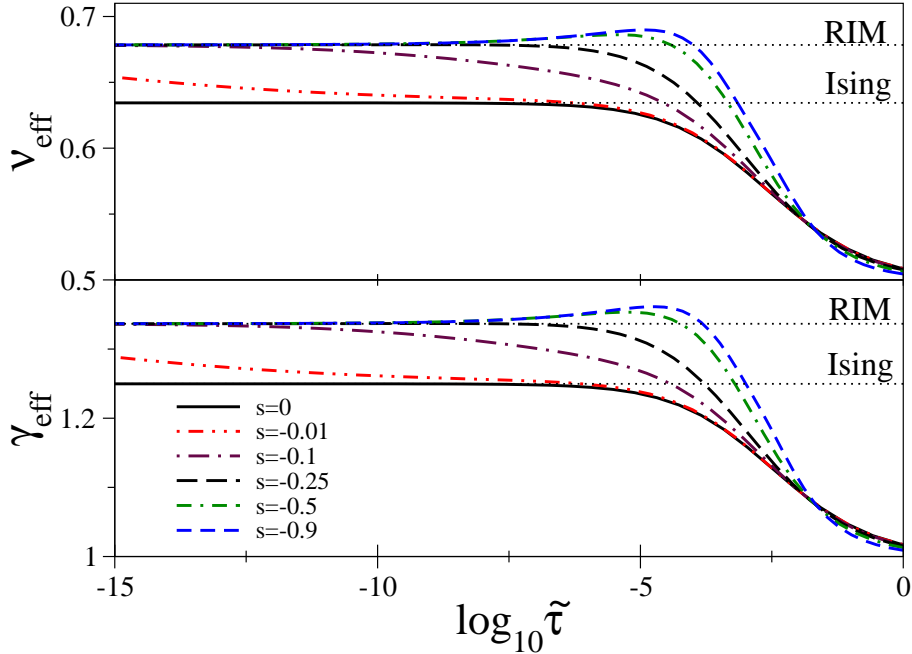}}
\vspace{2mm}
\caption{
Ising systems: The effective exponents $\gamma_{\rm eff}$ and
$\nu_{\rm eff}$ for several values of $s$ in the interval $-1<s<0$.
}
\label{fignueff}
\end{figure}

Using Eqs.~(\ref{flxi}), (\ref{flchi}), and (\ref{fltau}),
one can compute the crossover functions 
$\widetilde\xi$ and $\widetilde\chi$ along the RG trajectories, i.e.~for 
fixed $s$, and the corresponding effective exponents $\nu_{\rm eff}$
and $\gamma_{\rm eff}$, cf. Eq.~(\ref{effexp}).
The effective exponents 
$\gamma_{\rm eff}$ and $\nu_{\rm eff}$ are shown in Fig.~\ref{fignueff}
for several values of $s<0$ within the attraction domain of the
RIM FP. We note that they become nonmonotonic
for $s<s^*\approx -0.25$, where the RG trajectories reach the RIM FP
from below, see Fig.~\ref{figtraj}.

The crossover from the Gaussian FP to the 
RIM FP has also been investigated in Refs.~\cite{JOS-95,FHY-00}
in the framework of the minimal-subtraction scheme without 
$\epsilon$ expansion. The effective exponents computed here differ from those
defined in Refs.~\cite{JOS-95,FHY-00}, since there the nontrivial relation
between temperature and RG flow parameter was neglected. 
In spite of the different definitions, the crossover curves obtained
in Refs.~\cite{JOS-95,FHY-00} have the same qualitative features of those
shown in Fig.~\ref{fignueff}.

The above field-theoretical results may be related
with those obtained in a specific (lattice or experimental) system
by comparing the behavior in a neighborhood of the critical point.
Given a quantity ${\cal O}$, we can write for the field-theoretical
model
\begin{equation}
\langle {\cal O} \rangle \approx C(s) \tau^{-\sigma} 
   \left(1 + A_1(s) \tau^{\Delta_1} + A_2(s) \tau^{\Delta_2}\right),
\end{equation}
while for the lattice or experimental system we write
\begin{equation}
\langle {\cal O} \rangle \approx D \tau_s^{-\sigma}
   \left(1 + B_1 \tau_s^{\Delta_1} + B_2 \tau_s^{\Delta_2}\right).
\end{equation}
Then, we require these two expansions to agree apart from a rescaling of the 
reduced temperatures $\tau_s=c\tau$, i.e. 
\begin{equation}
B_1 = c^{\Delta_1} A_1(s), \qquad\qquad
B_2 = c^{\Delta_2} A_2(s),
\label{match-eq1}
\end{equation}
which gives
\begin{equation}
{A_1(s)\over B_1} = \left({A_2(s)\over B_2}\right)^{\Delta_1/\Delta_2}.
\label{match-eq2}
\end{equation}
Thus, in order to match the two expansions one should first determine $s$ 
by using Eq.~(\ref{match-eq2}) and then fix $c$ by using 
Eq.~(\ref{match-eq1}). This provides a mapping between the field-theoretical 
model and the considered system. This relation does not depend 
on the chosen quantity ${\cal O}$ because of the universality 
of ratios of subleading corrections (the ratios of the $B_1$'s and of the 
$B_2$'s of two different quantities are universal). 
Note, however, that the existence of this mapping is not
guaranteed. In particular, Eq.~(\ref{match-eq2}) requires 
${A_1(s)/B_1}$ and ${A_2(s)/B_2}$ to be both positive. Since $A_1(s)$
changes sign for $s = s^*$, it is always possible to have 
${A_1(s)/B_1} > 0$. But there is no guarantee that ${A_2(s)/B_2}$ 
can always be made positive. This is the well-known sign problem
that has been discussed at length in $O(M)$ models 
\cite{LF-90,Nickel-91,Schafer-94,Sokal-95}. For instance, it 
prevents to match the crossover curves for the scalar $\phi^4$ theory
with the results obtained for the three-dimensional Ising model. 
Ref.~\cite{Schafer-94} suggested the use of the ``strong-coupling" branch 
$g > g^*$, but this proposal fails in the massive zero-momentum
renormalization scheme  because of the 
nonanalyticity of the RG functions at the FP
\cite{Nickel,Sokal-95,PV-nonanal}. This phenomenon is even more 
evident in the RIM case, cf. Eq.~(\ref{v-at-RIM}).
It should also be stressed that the mapping defined by Eqs.
(\ref{match-eq1}) and (\ref{match-eq2}) does not imply that the
field-theoretical crossover curves exactly match the corresponding
ones for the considered system for all values of $\tau$.
In particular, there is no relation among the neglected coefficients
in the Wegner expansions.

Finally, let us discuss 
the RIM with nearest-neighbor interactions on a cubic lattice with 
spin density $p$. 
Numerical simulations show the presence of a dilution-independent 
continuous transition up to $p=0.40$ \cite{BFMMPR-98}. 
It is usually conjectured that the transition persists up to the 
percolation threshold of the spins $p=p_c$, 
$p_c=0.3116081(13)$ on a cubic lattice \cite{BFMMPR-99}. 
Below the percolation point the spins form finite domains and 
are therefore unable to show a critical behavior. 
It should be remarked that the transition for $p=p_c$ is not described by
the field-theory model (\ref{Hphi4rim}) and thus the RIM for $p=p_c$ 
does not correspond to $s= s_{\rm min}$. For the same reasons, the fact that
the transition disappears for $p<p_c$ does not provide evidence in favor of 
a finite $s_{\rm min}$. However, if the RIM can be related with  the 
field-theory model (for example if Eq.~(\ref{match-eq2}) can be solved for any
value of $p$) and the RIM with $p\to p_c$ corresponds 
to the field-theory model with $s\to \bar{s}$, then we can conclude 
$|\bar{s}| < |s_{\rm min}|$. Now we show that this condition 
is approximately verified. For this purpose, we must determine the 
relation between the RIM and the field-theory model. We use the 
results of Ref.~\cite{GL-76} that map the RIM
onto a translationally-invariant effective Hamiltonian
${\cal H}^{\rm RIM}_p$ for a field $\phi$. 
The expansion of ${\cal H}^{\rm RIM}_p$ for $\phi\to 0$ has the same form, 
up to order $\phi^4$, of the Hamiltonian
(\ref{Hphi4rim}) with $M=1$. The corresponding quartic couplings 
$u_0^{\rm RIM}$ and $v_0^{\rm RIM}$ appearing in this expansion 
are related to the magnetic concentration $p$ 
(note that such result does not depend
on the lattice type and on the spin-spin interaction as long as it is 
of short-range type) by
\begin{equation}
u_0^{\rm RIM}\propto p(p-1),\qquad v_0^{\rm RIM}\propto p, 
\end{equation}
and in particular 
\begin{equation}
{u_0^{\rm RIM} \over v_0^{\rm RIM}} = -{3\over 2}(1-p).
\label{sp}
\end{equation} 
It is tempting to assume $s \approx {u_0^{\rm RIM}/v_0^{\rm RIM}}$, 
which means that we neglect the fact that in ${\cal H}^{\rm RIM}_p$ 
there are interactions $\phi^n$ with any $n> 4$. 
The relation $s \approx - 3(1-p)/2$ follows. 
Using this relation and the numerical 
results of Refs.~\cite{BFMMPR-98,CMPV-03}, we can get an independent
approximate estimate of $s^*$. Since in the RIM on a cubic lattice 
one does not observe the leading scaling correction for 
$p^* \approx 0.8$, we obtain 
$s^* \approx - 0.3$, which is reasonably close to the 
FT estimate $s^* = - 0.25(5)$. 
Moreover, the percolation threshold $p_c$---$p_c=0.3116081(13)$ on a cubic 
lattice \cite{BFMMPR-99}---apparently corresponds 
to $\bar{s} \approx -1$, which is compatible with the predicted 
inequality $|\bar{s}|<|s_{\rm min}|$.

\subsection{Crossover from Ising to random critical behavior}
\label{crossIRIM}

The FT approach presented in Sec.~\ref{FTframework}
allows us to determine also the Ising-to-RIM crossover functions.
Considering in general a quantity ${\cal O}$ that behaves at the Ising 
FP, i.e.~in the absence of disorder, as $t^{-\rho_I \nu_I}$, 
standard RG arguments show that, in the limit $p\to 1$ and 
$t\equiv(T-T_I)/T_I\to 0$, where $T_I$ is the critical temperature
of the pure Ising model, ${\cal O}$ can be written in the scaling form
\begin{equation}
{\cal O} = a_0 t^{-\rho_I \nu_I} B_{\cal O}\left( g t^{-\phi} \right) =
a_1 \xi^{\rho_I} C_{\cal O}\left( g \xi^{\phi/\nu_I} \right),
\label{crf}
\end{equation}
where $g\propto 1-p$ is the scaling field
associated with disorder, which is a relevant perturbation of the Ising FP, 
and $a_0$ and $a_1$ are normalization constants.
The crossover exponent $\phi$ is equal \cite{Aharony-76} 
to the Ising specific-heat exponent $\alpha_I$, $\phi=\alpha_I$.
The functions $B_{\cal O}$ and $C_{\cal O}$ are universal, apart from 
trivial normalizations. By properly choosing $a_0$ and $a_1$ we can require 
$B_{\cal O}(0) = C_{\cal O}(0) = 1$. Another condition can be added by properly
fixing the normalization of $g$. 

\begin{figure}[tb]
\centerline{\psfig{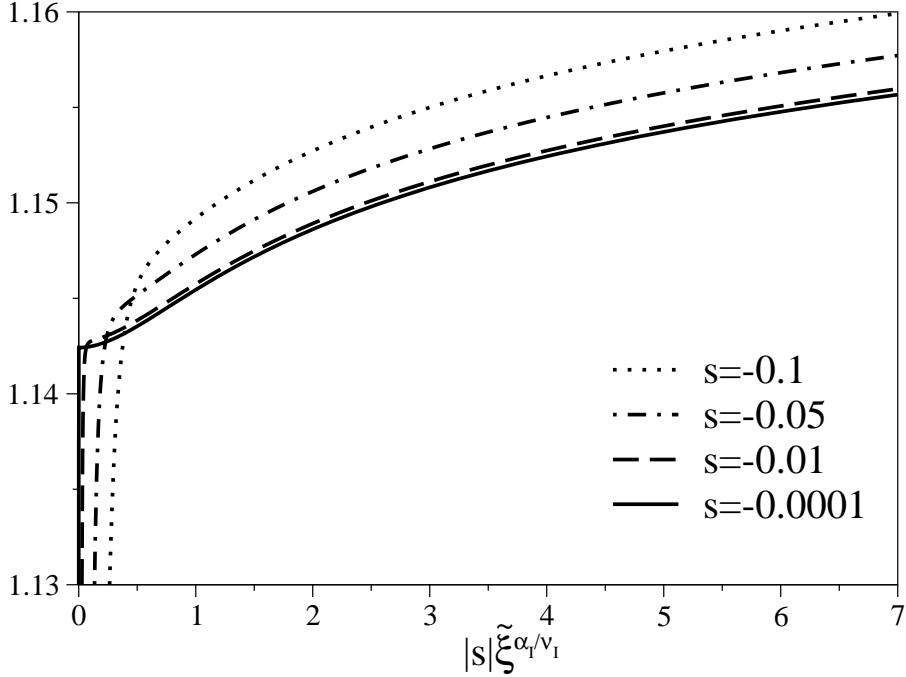}}
\vspace{2mm}
\caption{
Ising systems: 
The quantity $\widetilde\chi(\lambda,s)\lambda^{-2 + \eta_I}$
as a function of  $|s|\widetilde\xi^{\alpha_I/\nu_I}$ for several
values of $s$.
}
\label{scals}
\end{figure}

Within the FT approach the limit $g\to 0$ corresponds to $s\to 0^-$ and 
$ g \xi^{\phi/\nu_I} \sim s \lambda^{\phi/\nu_I}$. Therefore, 
crossover functions are obtained by taking the limit $s\rightarrow 0^-$ and
$\widetilde\xi=\lambda\rightarrow\infty$ of the quantity 
${\cal O} \xi^{-\rho_I}$, keeping $s\lambda^{\alpha_I/\nu_I}$ fixed.
In Fig.~\ref{scals} we show numerically that such a limit exists for the 
susceptibility $\chi$. We consider 
$\chi \xi^{-2 + \eta_I} = \widetilde\chi(\lambda,s) \lambda^{-2 + \eta_I}$
and plot this combination as a function of $|s|\widetilde\xi^{\alpha_I/\nu_I}$
for several values of $s$. The curves, obtained by using Eq.~(\ref{flchi})
and the conformal-mapping method, rapidly converge to a limiting function.

\begin{figure}[tb]
\centerline{\psfig{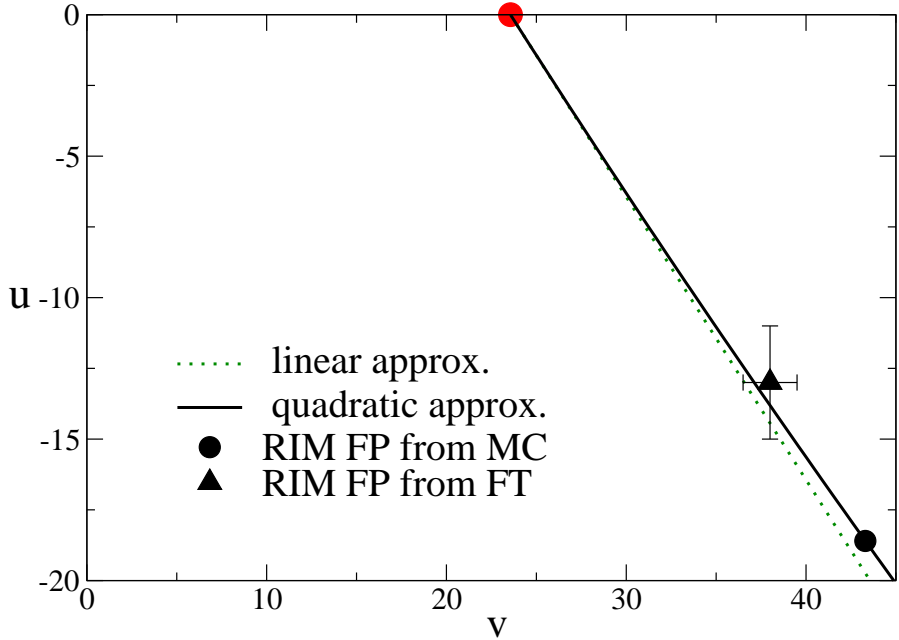}} 
\vspace{2mm}
\caption{
Plot of the curve $v=g(u)$. 
The full line represents the quadratic curve given in Eq.~(\ref{tu}), 
while the dotted line shows the linear approximation $g(u)=v_I-u$.
We also show the position of the RIM FP as obtained by
MC simulations of the RIM (circle) and by FT calculations (triangle).
}
\label{s0RG}
\end{figure}

In order to compute the crossover functions, we must first study the 
limit $s\rightarrow 0^-$ of the RG trajectories. As it can be seen from 
Fig.~\ref{figtraj}, in this limit the trajectory will eventually be formed 
by two parts connecting at the Ising FP: the line $u=0$ starting at the 
Gaussian FP and ending at the Ising FP, and a line $v = g(u)$ 
connecting the Ising FP to the RIM FP. The line $v=g(u)$ 
corresponds to a RG trajectory and therefore
$(u(\lambda),v(\lambda)) = [u(\lambda),g(u(\lambda))]$ must satisfy 
Eq.~(\ref{rgflow}). Therefore, $g(u)$ is the solution of the differential 
equation
\begin{equation}
{dg\over du} = {\beta_v(u,g(u))\over \beta_u(u,g(u))}
\label{eq-per-gu}
\end{equation}
with the initial condition $g(0) = v_I$. 
As discussed in App.~\ref{IsiRim},
$g(u)$ is expected to be analytic for $u\to 0$ and thus it can be 
expanded as 
\begin{equation}
g(u) = v_I + \sum_{n=1}^\infty g_n u^n.
\label{gu-exp}
\end{equation}
In App.~\ref{ap3} we compute the first coefficients: 
$g_1 = - 1$, a consequence of identity (\ref{identity}), 
$g_2 = 0.0033(1)$, and $g_3 = 1(2)\times 10^{-5}$. 

Since $g(u)$ corresponds to an RG trajectory with $s=0$, 
Eq.~(\ref{v-at-RIM}) implies that, close to the RIM FP, we have
\begin{equation}
g(u) \approx v^* + {1\over R_1}(u - u^*) - 
      u_{\lambda,2}(0) \left({1\over R_1} - {1\over R_2}\right)
      \left({u-u^*\over u_{\lambda,1}(0)}\right)^{\omega_2/\omega_1},
\label{gu-at-RIM}
\end{equation}
Eq.~(\ref{gu-at-RIM}) shows that $g(u)$ is not 
analytic at the RIM FP. Of course, one should check that $u_{\lambda,2}(0)$
does not vanish. We are not able to verify numerically this condition, but we 
believe that it is unlikely that $u_{\lambda,2}(0)=0$. Indeed, the curve 
$g(u)$ is a special curve only at the Ising FP, but it has no special status 
at the RIM FP and thus it should be nonanalytic as any generic RG trajectory 
\cite{Sokal-foot}.

The curve $g(u)$ can be computed \cite{differentmethod} 
by resumming the perturbative series
for the $\beta$ functions and then by explicitly solving Eq.~(\ref{eq-per-gu})
with the initial condition $g(0) = v_I$. 
The result turns out to be very well approximated by the simple 
expression
\begin{equation}
   g(u) \approx v_I - u + g_2 u^2,
\label{tu}
\end{equation}
where $v_I=23.56(2)$ is the coordinate of the
Ising FP \cite{CPRV-02} and $g_2\approx 0.0033$.
Such an approximation is effective, within the resummation errors,
up to the RIM FP. 
A graph is reported in Fig.~\ref{s0RG}.  
The results obtained by using 
[3,1], [4,1] and [5,1] Pad\'e-Borel approximants
would not be distinguishable from the curve (\ref{tu}) shown in Fig.~\ref{s0RG}.
For instance, $g(-13) \approx 37.1$ and $g(-18.6) \approx 43.3$, 
so that Eq.~(\ref{tu}) is perfectly compatible with the 
MC estimate of the FP, $u^* = -18.6(3)$, $v^* = 43.3(2)$, and with the
FT result, $u^* = -13(2)$, $v^* = 38.0(1.5)$, see Fig.~\ref{s0RG}.
The fact that both estimates lie on the limiting curve $v=g(u)$ shows that the 
FT approach is effective in determining the Ising-to-RIM trajectory, 
although it is apparently unable to determine precisely the position of the 
FP on this curve. As a final check, we compute $g'(u^*)$. Using 
Eq.~(\ref{gu-at-RIM}) and the estimate of $R_1$ reported in 
Sec.~\ref{resisi}, $R_1 = -0.90(2)$, we obtain 
$g'(u^*) = -1.11(2)$, while  Eq.~(\ref{tu}) gives
$g'(u^*) = -1.12$ (resp. $-1.09$) at the Monte Carlo estimate
(resp. field-theoretical) of the FP. 
The agreement is satisfactory.

Once we have determined $g(u)$, we can compute $u(\lambda,s)$ in the 
crossover limit. In App.~\ref{ap1}, we show that, in the crossover limit, 
$s\rightarrow 0^-$  keeping $|s|\lambda^{\alpha_I/\nu_I}$ fixed,
$u(\lambda,s)$ converges 
to $U(\sigma)$ which is implicitly defined by
\begin{equation}
\sigma = U(\sigma) \exp\left\{
      -{\alpha_I\over \nu_I} \int_0^{U(\sigma)}dx\; 
     \left[{1\over \beta_u(x,g(x))} + {\nu_I\over \alpha_I x}\right]\right\},
\label{Usigma}
\end{equation}
where 
\begin{equation}
\sigma \equiv 
    s \Sigma_1 \left({\lambda \over \Sigma_2}\right)^{\alpha_I/\nu_I},
\label{defsigma}
\end{equation}
and $\Sigma_1$ and $\Sigma_2$ are normalization constants such that 
$U(\sigma)\approx \sigma$ for $\sigma \to 0$. Their explicit expressions
are reported in App.~\ref{ap1}. Of course, $v(\lambda,s) = g(U(\sigma))$ 
in the scaling limit $s\to 0^-$. 
The curve $g(u)$ and Eq.~(\ref{Usigma}) completely 
fix the relevant RG trajectory in the crossover limit.

The computation of the crossover functions is then completely straightforward.
We consider the RG function ${\cal O}(\lambda,s)$ associated with $\cal O$
and assume that it satisfies the RG equation
\begin{equation}
\lambda {d{\cal O}\over d\lambda} = \rho(u,v) {\cal O},
\label{eq-calO}
\end{equation}
where $\rho(u,v)$ is the corresponding RG function such that 
$\rho(0,v_I) = \rho_I$. The crossover limit is studied in detail in 
App.~\ref{ap2}. We find that the crossover function $C_{\cal O}(y)$ 
can be written as 
\begin{equation}
C_{\cal O}(y) = \exp\left[ - \int_0^{U(\sigma)}dx\, 
   {\rho(x,g(x)) - \rho_I\over \beta_u(x,g(x))}\right],
\end{equation}
where the relation between $y$ and $\sigma$ should be fixed by 
choosing an additional normalization condition.
      
\begin{figure}[tb]
\centerline{\psfig{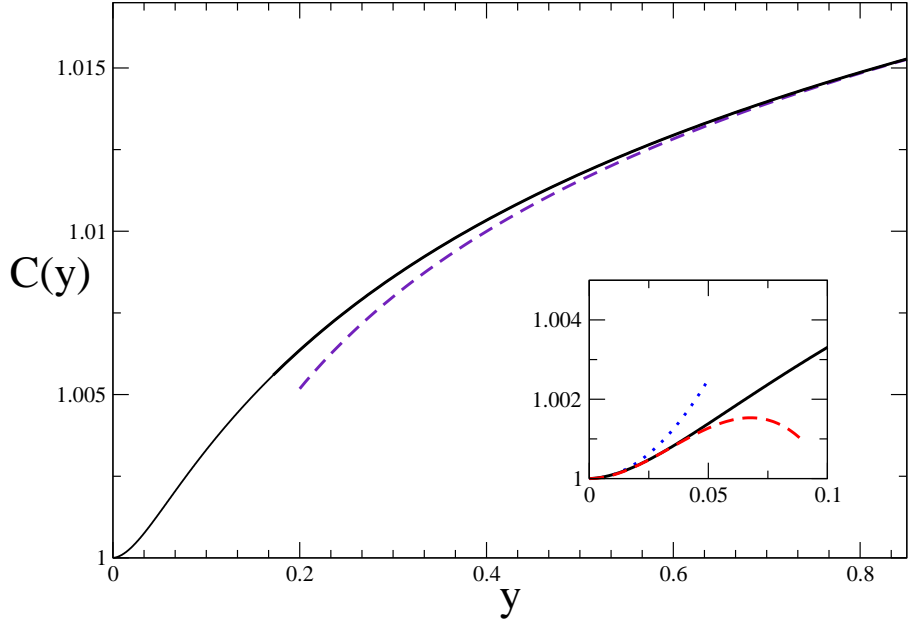}}
\vspace{2mm}
\caption{The crossover function $C_\chi(y)$ normalized 
according to Eq.~(\ref{nprmC}).
The dashed line represents the asymptotic behavior 
(\ref{cxasy}). The inset shows the small-$y$
behavior: the expansion to order $y^2$ corresponds to the 
dotted line while the expansion to order $y^3$ corresponds to the
dashed line. 
}
\label{crossfuncchi}
\end{figure}

We wish now to specialize the previous discussion to the magnetic 
susceptibility. In this case $\rho(u,v) = 2 - \eta_\phi(u,v)$. 
In order to completely specify the function $C_\chi(y)$ 
appearing in Eq.~(\ref{crf}) we must fix 
the normalization of $g$. We use the small-$y$ expansion of $C_\chi(y)$. 
Since $C_\chi'(0) = 0$, see App.~\ref{ap2}, we require 
\begin{equation}
   C_\chi(y) = 1 + y^2 + \sum_{n=3} c_n y^n,
\label{nprmC}
\end{equation}
for $y \to 0$ and $C_\chi(y)$ to be defined for $y > 0$. 
With these normalizations we have 
\begin{equation}
C_\chi(y) = \exp 
    \int_0^{U(\sigma)} dx\, {\eta_\phi(x,g(x))-\eta_I\over \beta_u(x,g(x))},
\label{Cchi}
\end{equation}
where $y = - y_0 \sigma$. The constant $y_0$ is positive
and is computed numerically in App.~\ref{ap3}:
$y_0 = 0.072(8)$.
The scaling function $C_\chi(y)$ is shown 
in Fig.~\ref{crossfuncchi}. 

We study the small-$y$ and large-$y$ behavior of $C_\chi(y)$. 
A rough estimate of the coefficient $c_3$ is $c_3=-4(2)$,
see App.~\ref{ap3}.
For large values of $y$, we have
\begin{equation}
C_\chi(y) \approx c_\infty y^{\nu_I (\eta_I-\eta)/\alpha_I},
\label{cxasy}
\end{equation}
where $\eta$ is the RIM exponent.
The best estimates of the exponents $\eta_I$ and $\eta$ 
of the Ising and RIM universality classes are 
$\eta_I=0.03639(15)$ (Ref.~\cite{CPRV-02}), 
$\eta_I=0.0368(2)$ (Ref.~\cite{DB-03}), and
$\eta = 0.035(2)$ (Ref.~\cite{CMPV-03}). These results suggest
$\eta_I>\eta$. This is confirmed by the analysis
of the fixed-dimension FT series: all analyses find 
$\eta_I>\eta$. In particular, analyses based on an expansion around the 
Ising FP \cite{CMPV-03} find $\eta_I-\eta=0.002(2)$.
This suggests that
$C_\chi(y)$ diverges for large $y$ with a very small exponent,
${\nu_I (\eta_I-\eta)/\alpha_I}=0.01(1)$.
We also estimated the coefficient $c_\infty$ appearing in the
large-$y$ behavior of $C_\chi(y)$, obtaining $c_\infty=1.05(5)$.
We proceeded as follows.
First, for given approximants of the RG functions,
we computed the exponents $\eta_I$, $\nu_I$, and 
$\eta$, and the function $C_\chi(y)$. Then, we calculated
$C_\chi(y) y^{-\nu_I (\eta_I-\eta)/\alpha_I}$ and determined the constant 
$c_\infty$ from its large-$y$ behavior. This procedure gave an estimate of 
$c_\infty$ for a given set of approximants. The final result was obtained as 
usual, by comparing the results of different approximants and of series of 
different order.

\subsection{Crossover in randomly dilute multicomponent spin systems}
\label{multisec}

\begin{figure}[tb]
\centerline{\psfig{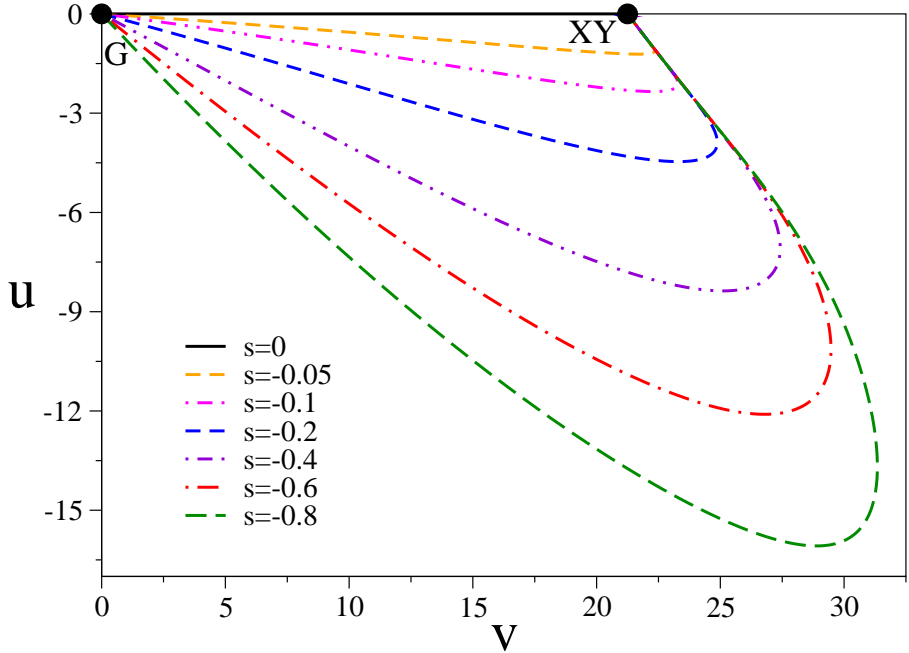}}
\vspace{2mm}
\caption{
RG trajectories in the dilute XY model for $-1<s\leq 0$.
}
\label{figtraj2}
\end{figure}

\begin{figure}[tb]
\centerline{\psfig{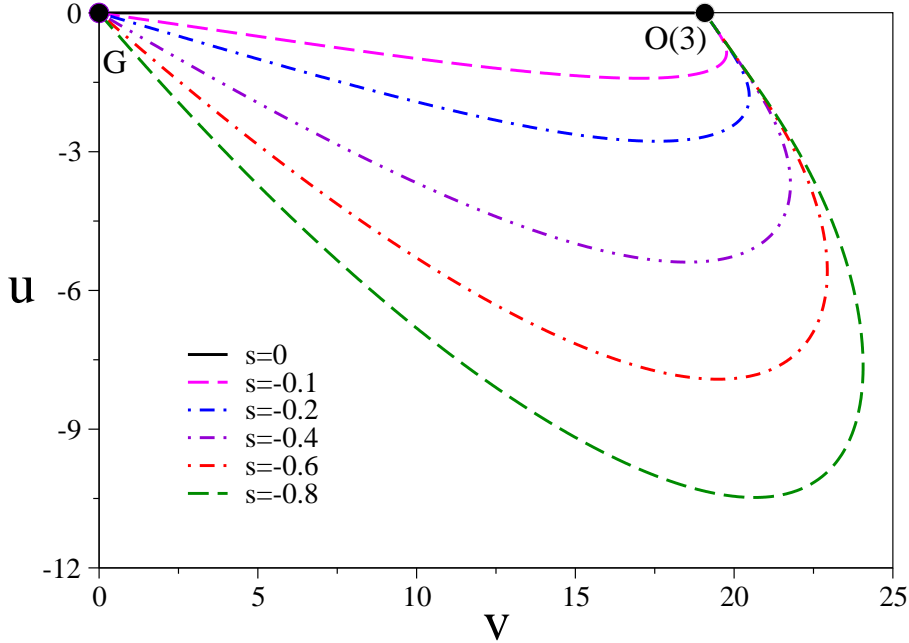}}
\vspace{2mm}
\caption{
RG trajectories in the dilute Heisenberg model for $-1<s\leq 0$.
}
\label{figtraj3}
\end{figure}

In the case of multicomponent systems, the stable FP 
is the O($M$)-symmetric FP $(0,v^*)$.
Precise estimates of $v^*$ 
have been obtained by employing FT and lattice techniques 
\cite{PV-r,CHPRV-01,CHPRV-02,PV-00-g,GZ-98}:
$v^*=21.16(5)$ (FT) and $v^*=21.14(6)$ (lattice)
for the XY universality class,
$v^*=19.06(5)$ (FT) and $v^*=19.13(10)$ (lattice)
for the Heisenberg universality class.

\begin{figure}[tb]
\centerline{\psfig{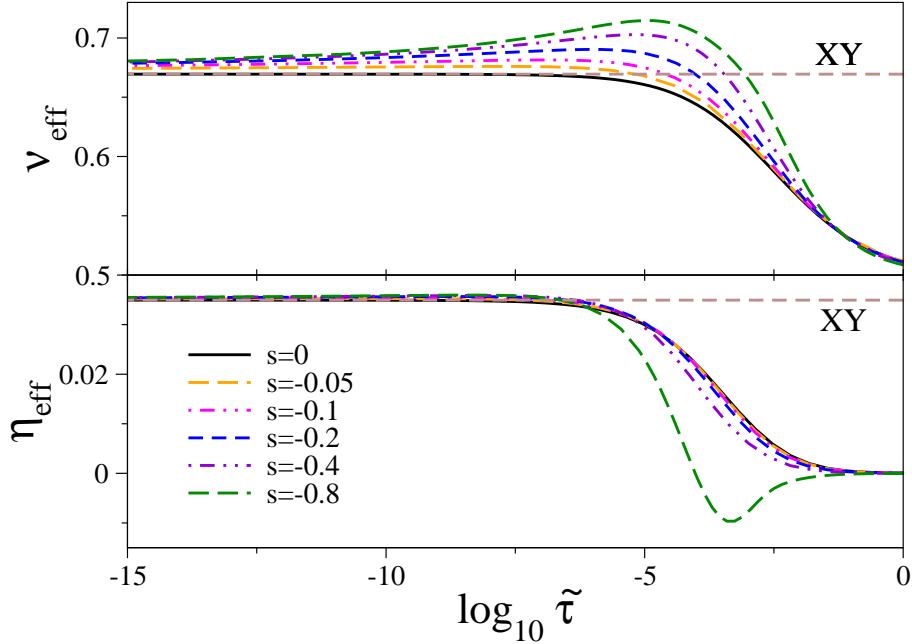}}
\vspace{2mm}
\caption{
The effective exponents $\eta_{\rm eff}$ and
$\nu_{\rm eff}$ of the dilute XY model for $-1<s\leq 0$.
In the Gaussian limit $\nu = 1/2$ and $\eta=0$, while in the 
Wilson-Fisher limit $\nu = 0.67155(27)$ and $\eta = 0.0380(4)$ 
(Ref.~\protect\cite{CHPRV-01}).
}
\label{fignueff2}
\end{figure}

\begin{figure}[tb]
\centerline{\psfig{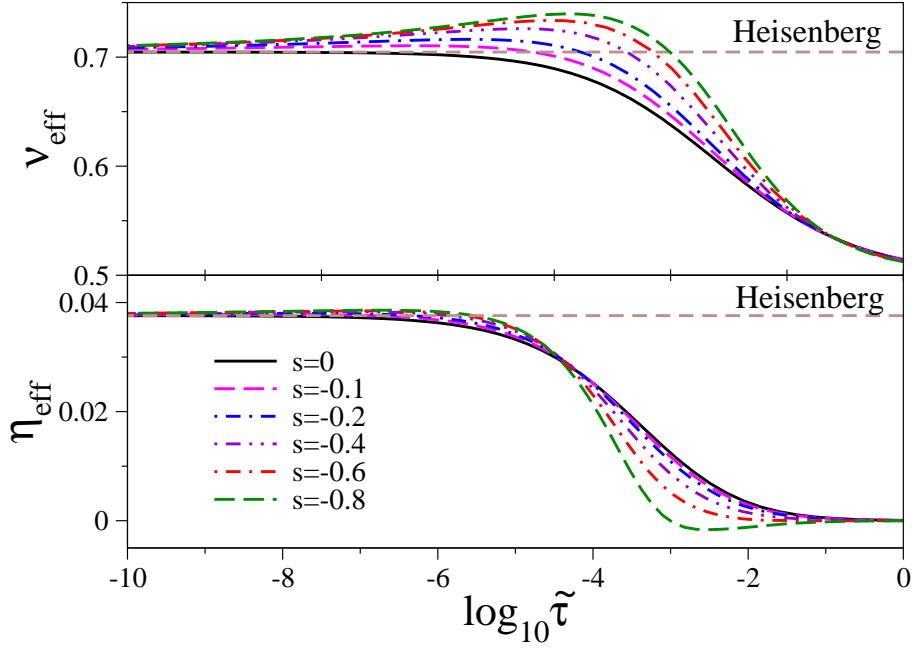}}
\vspace{2mm}
\caption{
The effective exponents $\eta_{\rm eff}$ and
$\nu_{\rm eff}$ of the dilute Heisenberg model for $-1<s\leq 0$.
In the Gaussian limit $\nu = 1/2$ and $\eta=0$, while in the
Wilson-Fisher limit $\nu = 0.7112(5)$ and $\eta = 0.0375(5)$
(Ref.~\protect\cite{CHPRV-02}).
}
\label{fignueff3}
\end{figure}

In Figs.~\ref{figtraj2} and \ref{figtraj3} we show, 
respectively for XY and Heisenberg systems,
the RG trajectories in the $u,v$ plane for several values of $s$ in the range
$-1<s\leq 0$. Figs.~\ref{fignueff2} and \ref{fignueff3}
report the corresponding effective exponents $\eta_{\rm eff}$ and 
$\nu_{\rm eff}$ respectively for XY and Heisenberg systems.
They are nonmonotonic.
In particular, for $s$ close to $-1$, $\eta_{\rm eff}$ becomes 
negative for intermediate values of $\tilde{\tau}$.
As in the Ising case, the resummations become
less reliables---and again
hint at runaway trajectories---for $s\lesssim -1$.

Finally, we mention that the RG trajectories and the effective 
crossover exponents of dilute Heisenberg systems
have been recently investigated in Ref.~\cite{DFHI-03},
using a two-loop approximation within the minimal-subtraction
scheme without $\epsilon$ expansion and neglecting the 
nontrivial relation between temperature and RG flow parameter.
In spite of all simplifying assumptions,
the results are in qualitative agreement with ours.
Moreover, Ref.~\cite{DFHI-03} discusses crossover phenomena
observed in experiments on isotropic magnets, showing several results for the
effective exponents that are in qualitative agreement with the
curves shown in Fig.~\ref{fignueff3}.

\acknowledgments

We thank Aleksandr Sokolov 
for useful and interesting discussions.

\appendix

\section{Some relations among the RG functions}
\label{identities}

In this Section we prove identities (\ref{identity}) and 
(\ref{etaidentity}) holding along the $u=0$ axis, and (\ref{identityv}), 
(\ref{etaidentityv}), and (\ref{etatidentityv}) holding along the $v=0$ axis.

Let us first prove the identities along the $u=0$ axis
in the case $M=1$; the extension to other values of $M$ is straightforward.
We consider a generic theory with fields $\phi^A$ and Hamiltonian density
\begin{equation}
{\cal H} = {1\over2} \sum_{\mu,A} (\partial_\mu \phi^A)^2 + 
           {1\over2} r \sum_A (\phi^A)^2 + 
           {g\over 4!} \sum_A (\phi^A)^4 + 
           {1\over 4!} \sum_{ABCD} C_{ABCD} \phi^A \phi^B \phi^C \phi^D.
\end{equation}
For $C_{ABCD} = 0$ the model is simply a collection of decoupled 
Ising $\phi^4$ theories. In order to compute the corrections to first 
order in $C_{ABCD}$, we consider the one-particle irreducible correlation 
functions of the fields 
expressed in terms of the bare couplings $g$ and $C_{ABCD}$ and 
of the inverse susceptibility $\chi^{-1}$ as effective mass
(the results also hold for the massless theory in dimensional regularization)
\begin{equation}
\Gamma_{A_1,\ldots,A_n} = 
  \langle \phi^{A_1},\ldots, \phi^{A_n}\rangle.
\end{equation}
Then, we prove that, if all indices are equal, 
\begin{equation}
\Gamma_{A,A, \ldots, A} = f(g) + C_{AAAA} {\partial f(g) \over \partial g} + 
      O(C^2) = f(g + C_{AAAA}) + O(C^2).
\label{relationGamma}
\end{equation}
Using this relation, one can derive identities (\ref{identity}) and 
(\ref{etaidentity}). Indeed, Eq.~(\ref{relationGamma})
 implies that (setting $\bar{u}_0 \equiv u_0/m$ and
$\bar{v}_0 \equiv v_0/m$)
\begin{eqnarray}
&& Z_\phi = f_\phi(\bar{u}_0 + \bar{v}_0) + O(\bar{v}_0^2),\\
&& u+v = f_{u+v}(\bar{u}_0+\bar{v}_0) + O(\bar{v}_0^2),\\
&& \beta_u+\beta_v = f_\beta(\bar{u}_0+\bar{v}_0) + O(\bar{v}_0^2).
\end{eqnarray}
To prove Eq.~(\ref{relationGamma}), consider a generic diagram $D$ 
contributing to the correlation function. If $\chi^{-1}$ is used as effective
mass or the mass vanishes and dimensional regularization is used, 
the diagram has the following properties: it does not contain 
tadpole subgraphs; given a vertex $V$, the subdiagram $D/V$ obtained 
by deleting the lines going out of $V$ may be disconnected, but each 
piece contains at least one external line. The contribution of the 
diagram $D$ is the product of three factors: 
the first is the integral over the 
internal momenta, the second the symmetry factor, and the third one---we call
it $I(D)_{A,A,\ldots,A}$---takes 
into account the interaction vertex 
\begin{equation}
   V_{ABCD} = - g \delta_{ABCD} - C_{ABCD}.
\end{equation}
Clearly, we are only interested in the last term which can be written in the 
form 
\begin{eqnarray}
   I(D)_{A,A,\ldots,A} &=& \left. I(D)_{A,A,\ldots,A}\right|_{C=0} - 
   \sum_{V\in D} 
            \left.  I(D/V)_{A,A,\ldots,A;I,J,K,L}\right|_{C=0}  C_{IJKL} 
\nonumber \\ 
   &=& 
    (-g)^n - n (-g)^{n-1} C_{AAAA},
\label{eq2}
\end{eqnarray}
where $n$ is the number of vertices of $D$. In the last step, we have used the 
two properties we have mentioned above: they guarantee that 
$I(D/V)_{A,A,\ldots,A;I,J,K,L} = (-g)^{n-1} \delta_{AIJKL}$, since for $C=0$
a connected diagram does not vanish only if the indices on the external legs 
are all equal. Eq.~(\ref{eq2}) gives immediately Eq.~(\ref{relationGamma}).
   
Identities (\ref{identityv}), (\ref{etaidentityv}), 
and (\ref{etatidentityv}) along the $v=0$ axis
can be proved in a similar fashion. 
Let us again restrict ourselves to the case $M=1$, the 
extension to generic values of $M$ being straightforward. 
Consider the Hamiltonian density
\begin{equation}
{\cal H} = {1\over2} \sum_{\mu,A} (\partial_\mu \phi^A)^2 + 
           {1\over2} r \sum_A (\phi^A)^2 + 
           {g\over 4!} \sum_{AB} (\phi^A)^2 (\phi^B)^2+ 
           {1\over 4!} \sum_{ABCD} C_{ABCD} \phi^A \phi^B \phi^C \phi^D,
\end{equation}
where $C_{ABCD}$ is symmetric in all indices.
For $C_{ABCD} = 0$ the model is simply an $N$-vector $\phi^4$ theory,
where $N$ is the dimension of the field.
In order to compute the corrections to first order in $C_{ABCD}$, 
we consider here a different set of correlation functions:
$O(N)$-invariant (therefore there are no external indices) 
one-particle irreducible correlation 
functions of the fields and of any $O(N)$-invariant operator. 
Consider again a diagram $D$, a vertex $V$, and the 
interaction contribution $I(D/V)_{I,J,K,L}$ for $C=0$. Because of the 
$O(N)$ invariance, its symmetrized part is given by
\begin{equation}
\left. I(D/V)_{\{I,J,K,L\} }\right|_{C=0}  = 
        \hat{I}(D/V) (\delta_{IJ}\delta_{KL} + 
       \delta_{IK}\delta_{JL} + \delta_{IL}\delta_{JK})\; .
\end{equation}
Then, repeating the argument leading to Eq.~(\ref{eq2}), we obtain
\begin{equation}
I(D) = \left. I(D)\right|_{C=0} - 3 \sum_V \hat{I}(D/V) \sum_{IJ} C_{IIJJ} .
\end{equation} 
The constant $\sum_V \hat{I}(D/V)$ is determined by 
computing the derivative of $I(D)$ with respect to $g$ at $C = 0$. 
\begin{eqnarray}
\left. {\partial I(D)\over \partial g} \right|_{C=0} &=&
   \sum_{V\in D} \sum_{IJKL} \left. I(D/V)_{I,J,K,L}\right|_{C=0} \times 
   \left(-{1\over3}\right) (\delta_{IJ}\delta_{KL} +
       \delta_{IK}\delta_{JL} + \delta_{IL}\delta_{JK}) 
\nonumber \\ 
   &= &
   - \sum_{V\in D} \hat{I}(D/V) N(N+2)\; .
\end{eqnarray}
It follows
\begin{equation} 
I(D) = f\left(g + {3 \sum_{IJ} C_{IIJJ}\over N(N+2)}\right) + O(C^2),
\label{id}
\end{equation} 
where $f(g) = \left. I(D)\right|_{C=0}$. This relation is valid only for 
$O(N)$-invariant quantities, but it can also be applied to the correlation 
functions of the elementary fields by simply contracting the external indices. 
It allows us to derive a number of relations involving the
$\beta$-functions and the RG functions associated with the exponents.
For example, considering the 
$MN$ model (\ref{Hphi4rim}) for $M=1$, relation (\ref{id}) implies
Eqs.~(\ref{identityv}), (\ref{etaidentityv}),
and (\ref{etatidentityv}) with  $M=1$.

\section{The Ising-to-RIM crossover}
\label{IsiRim}

In this appendix we compute the limit $s\rightarrow 0^-$ of the RG
trajectories and the Ising-to-RIM crossover function $C_{\cal O}(y)$, 
cf. Eq.~(\ref{crf}).

\subsection{The limit $s\rightarrow 0^-$ of the RG trajectories}
\label{ap1}   

Here, we wish to prove Eqs.~(\ref{Usigma}) and (\ref{defsigma}) that 
give $u(\lambda,s)$ in the crossover limit
$s\rightarrow 0^-$  keeping $|s|\lambda^{\alpha_I/\nu_I}$ fixed.
As discussed in Sec.~\ref{crossIRIM}, in the crossover limit
the RG trajectory is formed by two parts connecting at the Ising FP: 
the line $u=0$ starting at the
Gaussian FP and ending at the Ising FP, and the line $v = g(u)$
connecting the Ising FP to the RIM FP. Now, we will solve the flow equations
(\ref{rgflow}) in the two cases and we will match the two solutions in the 
neighborhood of the Ising FP. 
Let us consider first the behavior near $v = g(u)$. The flow equation for 
$u(\lambda,s)$ can be written as 
\begin{equation}
- \lambda {du\over d\lambda} = \beta_u(u,g(u)).
\end{equation}
Since $\beta_u(u,v_I) = -u\alpha_I/\nu_I$ for $u\to 0$, we can write
the solution as 
\begin{equation}
\lambda = A(s) u(\lambda,s)^{\nu_I/\alpha_I} 
   \exp\left[-\int_0^{u(\lambda,s)} dx\, 
       \left({1\over \beta_u(x,g(x))} + {\nu_I\over\alpha_I x}\right)
       \right],
\label{lambda-u-cross}
\end{equation}
where $A(s)$ is a (at this stage unknown) function of $s$. 

Now let us consider the second case, i.e. the trajectory near the $u=0$ axis.
For $u\to 0$, we can write $\beta_u(u,v) = u f(v) + O(u^2)$, with 
$f(0) = -1$, $f(v_I) = -\alpha_I/\nu_I$. As for $\beta_v(u,v)$ we simply 
set $u=0$. Note that $\beta_v(0,v) = -v + O(v^2)$ for $v\to 0$ and 
$\beta_v(0,v) = - \omega_I (v_I - v)$ for $v\to v_I$. In the limit
we are interested in, the RG equations (\ref{rgflow}) become
\begin{equation}
- \lambda {dv\over d\lambda} = \beta_v(0,v), \qquad\quad
- \lambda {du\over d\lambda} = u f(v).
\end{equation}
Keeping into account the initial conditions (\ref{ini-rgflow}), we obtain 
\begin{eqnarray}
\lambda &=& v \exp\left[- \int_0^v dx\, \left(
       {1\over \beta_v(0,x)} + {1\over x}\right)\right], 
\label{lambda-v} \\
u &=& sv \exp \int_0^v dx\, \left(
       {f(x)\over \beta_v(0,x)} - {1\over x}\right).
\label{u-v} 
\end{eqnarray}
Eqs.~(\ref{lambda-v}) and (\ref{u-v}) implicitly define $u(\lambda,s)$. 
We must now match the two solutions near the Ising FP, determining 
the unknown constant $A(s)$.
If we define
\begin{eqnarray}
\Sigma_1 &\equiv& v_I \exp \int_0^{v_I} dx\, \left(
       {f(x)\over \beta_v(0,x)} - {1\over x} - 
       {\alpha_I\over \nu_I \omega_I (v_I - x)} \right),
\nonumber  \\
\Sigma_2 &\equiv& v_I \exp\left[- \int_0^{v_I} dx\, \left(
       {1\over \beta_v(0,x)} + {1\over x} + 
       {1\over \omega_I (v_I - x)} \right)\right],
\end{eqnarray}
for $v\to v_I$ Eqs.~(\ref{lambda-v}) and (\ref{u-v}) can be written as 
\begin{eqnarray}
\lambda &\approx& \Sigma_2 (v_I - v)^{-1/\omega_I}, \nonumber \\
u &\approx& s \Sigma_1 (v_I - v)^{-\alpha_I/ \nu_I \omega_I}.
\label{ulambda-Deltav}
\end{eqnarray}
Therefore, for $v\to v_I$ we have
\begin{equation}
u(\lambda,s) \approx
 s \Sigma_1 \left(\lambda\over\Sigma_2\right)^{\alpha_I/ \nu_I}.
\label{eq1}
\end{equation}
On the other hand, Eq.~(\ref{lambda-u-cross}) gives for $u\to 0$,
\begin{equation}
u(\lambda,s) \approx (\lambda/A(s))^{\alpha_I/ \nu_I}.
\label{eq1b}
\end{equation}
By comparing Eqs.~(\ref{eq1}) and (\ref{eq1b}) we obtain $A(s)$. 
Finally, Eq.~(\ref{lambda-u-cross}) can be written as 
\begin{equation}
s \Sigma_1 \left(\lambda\over\Sigma_2\right)^{\alpha_I/ \nu_I} = 
   u(\lambda,s) \exp\left[-{\alpha_I\over \nu_I} \int_0^{u(\lambda,s)} dx\, 
       \left({1\over \beta_u(x,g(x))} + {\nu_I\over\alpha_I x}\right)
       \right].
\label{eq1c}
\end{equation}
This ends the proof of Eqs.~(\ref{Usigma}) and (\ref{defsigma}).
   
\subsection{Crossover functions}
\label{ap2}

The computation of the crossover function is similar to that 
presented in App.~\ref{ap1}. 
We first consider the RG equation (\ref{eq-calO}) on the line $v = g(u)$.
Using the flow equation for $u(\lambda,s)$ we can write
\begin{equation}
{d{\cal O}\over du} = - {\rho(u,g(u))\over \beta_u(u,g(u))} {\cal O}.
\end{equation}
The solution can be written as 
\begin{equation}
{\cal O} = B(s) u(\lambda,s)^{\rho_I\nu_I/\alpha_I}
   \exp\left[-\int_0^{u(\lambda,s)} dx\,
       \left({\rho(x,g(x))\over \beta_u(x,g(x))} + 
             {\rho_I\nu_I\over\alpha_I x}\right)
       \right],
\label{calO-u-cross}
\end{equation}
where $B(s)$ is an unknown function.

For $u\to 0$, we can use the flow equation for $v(\lambda,s)$ and write
\begin{equation}
{d{\cal O}\over dv} = - {\rho(0,v)\over \beta_v(0,v)} {\cal O}.
\end{equation} 
We assume $\rho(0,0) = \rho_0$ ($\rho_0$ is the naive Gaussian dimension of 
${\cal O}$) and ${\cal O}\approx {\cal O}_0 \lambda^{\rho_0}$ at the 
Gaussian FP (${\cal O}_0$ is a normalization constant). Then, the
previous equation gives
\begin{equation} 
{\cal O}= {\cal O}_0 v^{\rho_0} \exp \left[ - \int_0^v dx\, \left(
       {\rho(0,x)\over \beta_v(0,x)} + {\rho_0\over x}\right)\right].
\end{equation}
Now, we must compute the behavior for $v\to v_I$. Defining
\begin{equation} 
T_1 \equiv {\cal O}_0 v^{\rho_0}_I \exp \left[ - \int_0^{v_I} dx\, \left(
       {\rho(0,x)\over \beta_v(0,x)} + {\rho_0\over x} + 
       {\rho_I\over \omega_I (v_I - x)} \right)\right],
\end{equation}
we obtain for $v\to v_I$
\begin{equation}
{\cal O} \approx T_1 (v_I - v)^{-\rho_I/\omega_I}
         \approx T_1 \left(\lambda\over\Sigma_2\right)^{\rho_I}
         \approx T_1 \left({u\over s \Sigma_1}\right)^{\nu_I\rho_I/\alpha_I},
\end{equation}
where we have used Eq.~(\ref{ulambda-Deltav}).
On the other hand, Eq.~(\ref{calO-u-cross}) gives in the limit $u\to 0$
\begin{equation}
{\cal O} \approx B(s) u^{\nu_I\rho_I/\alpha_I}.
\end{equation}
Therefore,
\begin{equation}
{\cal O} = T_1 \left({u\over s \Sigma_1}\right)^{\nu_I\rho_I/\alpha_I}
   \exp\left[-\int_0^{u(\lambda,s)} dx\,
       \left({\rho(x,g(x))\over \beta_u(x,g(x))} + 
             {\rho_I\nu_I\over\alpha_I x}\right)
       \right].
\label{calO-u-cross2}
\end{equation}
Finally, by using Eq.~(\ref{eq1c}) to eliminate $u^{\nu_I\rho_I/\alpha_I}$,
we obtain 
\begin{equation}
{\cal O} = T_1 \Sigma_2^{-\rho_I} \lambda^{\rho_I}
\exp\left[- \int_0^{U(\sigma)} dx\;
   {\rho(x,g(x)) - \rho_I\over \beta_u(x,g(x))}\right].
\end{equation}
The crossover function $C_{\cal O}(y)$ normalized so that $C_{\cal O}(0) = 1$
is then given by
\begin{equation}
C_{\cal O}(y) = \exp \left[ - \int_0^{U(\sigma)} dx\; 
   {\rho(x,g(x)) - \rho_I\over \beta_u(x,g(x))}\right].
\label{CcalO-gen}
\end{equation}
To fully specify the function $C_{\cal O}(y)$ we must also relate $y$ with 
$\sigma$ by adding an additional normalization condition. For the magnetic
susceptibility this is done in detail in Sec.~\ref{crossIRIM}. 

We can specialize these results to the observables we have considered in the
paper. First, we consider the four-point quartic couplings $G_{22}$ and 
$G_4$. Since they are related to $u$ and $v$, $G_{22}= {u/3}$ and $G_4=v$
(see Ref.~\cite{CPV-03-eqst}), and 
$u\approx U(\sigma)$, $v\approx g(u)$ in the crossover limit, we obtain
\begin{eqnarray}
&& C_{G_{22}}(y) = {U(\sigma)\over \sigma}, \\
&& C_{G_4}(y) = {1\over v_I} g(U(\sigma)).
\end{eqnarray}
Note that $C_{G_{22}}(y)$ is not simply $U(\sigma)$ since the crossover 
function is defined by $u\sim \lambda^{\alpha_I/\nu_I} C_{G_{22}}(y)$. 
These equations can also be derived from Eq.~(\ref{CcalO-gen}) bu using 
$\rho(u,v) = -\beta_u(u,v)/u$ and $\rho(u,v) = -\beta_v(u,v)/v$ for 
$u$ and $v$ respectively.

Finally, let us consider the magnetic susceptibility $\chi$. In this case
$\rho(u,v) = 2 - \eta_\phi(u,v)$. Thus, by using Eq.~(\ref{CcalO-gen}) we 
obtain Eq.~(\ref{Cchi}). Let us now show that ${C}_\chi'(0) = 0$. First, 
note that, because of identity (\ref{etaidentity}), near the Ising FP we have
\begin{equation}
\eta_\phi(u,v) - \eta_I = A(u+v-v_I) [1 + O(u) + O(v-v_I)],
\end{equation}
where $A$ is a constant. Then, since $g(u) = v_I - u + O(u^2)$, we obtain
$\eta_\phi(u,g(u)) - \eta_I = O(u^2)$. Substituting in Eq.~(\ref{Cchi}),
this gives immediately $C_\chi'(0) = 0$.

Finally, we argue that the crossover function
$C_{\cal O}(y)$ and $g(u)$ (that can be related to the crossover function
of $v = G_4$) are analytic for $y\to 0$ and $u\to 0$ respectively.
This is not obvious since for $u=0$ RG functions are nonanalytic at the 
Ising FP \cite{Nickel,PV-nonanal}. We will now show that such a problem
does not arise for the RG functions defined along the crossover line 
$v = g(u)$. The reason is that such a line has a very special status at the 
Ising FP: It is the line that is tangent to the relevant direction associated
with disorder and that is {\em orthogonal} to all irrelevant directions.

To clarify the issue, let us for instance consider the singular part of the 
free energy. In a neighborhood of the Ising FP it can be written as 
\cite{Wegner-76} 
\begin{equation}
F_{\rm sing} = f_t^{d/\nu_I} 
    F\left(f_p f_t^{-\phi},\{f_i f_t^{\Delta_i}\}\right),
\end{equation}
where $f_t$, $f_p$, and $\{f_i\}$ are the nonlinear scaling fields associated
with the temperature, the dilution, and the irrelevant RG operators.
For $t\equiv (T - T_I)/T_I\to 0$ and $p\to 1$, $f_t \sim t$ and 
$f_p \sim (1-p) \sim g$. The exponents $\Delta_i$ are associated with the 
irrelevant operators and are positive. A basic result of RG theory is 
that the nonlinear scaling fields are analytic in $t$ and $p$ and 
the function $F$ is analytic in all its arguments. In the crossover
limit, $f_i$ approaches a constant and $f_t$ goes
to zero, so that $f_i f_t^{\Delta_i}\to 0$. It follows
\begin{equation}
F_{\rm sing} \approx t^{d/\nu_I} F(gt^{-\phi}, \{0\}),
\end{equation}
which shows that the crossover function associated with $F_{\rm sing}$ is 
analytic in $gt^{-\phi}$. The argument can be trivially generalized to any
zero-momentum quantity; we conjecture that it also applies to quantities
involving the correlation length.

\subsection{Some numerical results}
\label{ap3}

In this Section we report some details on the numerical computation of 
$g(u)$ and $C_\chi(y)$. 
Let us first focus on the determination of the coefficients $g_n$ defined in
Eq.~(\ref{gu-exp}). They have been obtained by resumming perturbative 
series $g_n(v)$ such that $g_n = g_n(v_I)$. For the purpose of determining 
$g_n(v)$, we write 
\begin{eqnarray}
\beta_u(u,v) &=& \sum_n b_{u,n}(v) u^n, \\
\beta_v(u,v) &=& \sum_n b_{v,n}(v) u^n.
\end{eqnarray}
Then, by using Eq.~(\ref{eq-per-gu}), we obtain
\begin{equation}
g_2 = \left. {b_{v,0}''(v) - 2 b_{v,2}(v) - 2 b_{u,2}(v) \over 
       2 [\alpha_I/\nu_I + b_{v,1}(v)]} \right|_{v = v_I},
\end{equation}
and similar, but more complex, expressions for $g_3$, $g_4$, etc. 
The series $g_n(v)$ can be obtained by expanding the right-hand side in powers
of $v$. For $g_2$ and $g_3$ we obtain
\begin{eqnarray}
g_2(\bar{v}) &=&  
0.00663146 - 0.00693165 \bar{v} + 0.0116887 \bar{v}^2  - 0.0225971\bar{v}^3 +
\label{g2ser}\\ 
&&+ 0.0455962 \bar{v}^4 - 0.0954011 \bar{v}^5 + O(\bar{v}^6),
\nonumber \\
g_3(\bar{v}) &=&  
0.0000293176 - 0.0000813454 \bar{v} + 0.000206937 \bar{v}^2  - 0.000485549 \bar{v}^3 +
\label{g3ser}\\ 
&& + 0.00110105 \bar{v}^4 + O(\bar{v}^5),
\nonumber
\end{eqnarray}
where $\bar{v}\equiv 3 v/(16\pi)$. By resumming these series we get
\begin{equation}
g_2=0.0033(1),\qquad g_3 = 1(2)\times 10^{-5}.
\end{equation}
We computed 
the function $g(u)$ by using Eq.~(\ref{eq-per-gu}), i.e.~without relying on an
expansion around the Ising FP, and by resumming the $\beta$-functions
using [3/1], [4/1], and [5/1] Pad\`e-Borel approximants
constrained to have a zero at $v=v_I=23.56$.
The results up to $u \approx - 20$ 
would not be distinguishable from the quadratic approximation
shown in Fig.~\ref{s0RG}.

Let us now consider $C_\chi(y)$. This function can be computed directly
by using Eqs.~(\ref{Usigma}) and (\ref{Cchi}). They provide 
$C_\chi$ as a function of the variable $\sigma$. In order to 
compute the relation between $\sigma$ and $y$, we need to determine 
the small-$\sigma$ behavior of $C_\chi$. We write 
\begin{equation}
C_\chi = 1 + \sum_{n=2} \overline{c}_n \sigma^n,
\end{equation}
and, as for $g(u)$, we compute perturbative series $\overline{c}_n(v)$
such that $\overline{c}_n = \overline{c}_n(v_I)$. By resumming these 
expansions we obtain
\begin{equation}
\overline{c}_2 = 0.0052(12), \qquad 
c_3 = \overline{c}_3 \overline{c}_2^{-3/2} = -4(2).
\end{equation}
The variable $y$ defined by the normalization condition (\ref{nprmC}) 
is related to $\sigma$ by 
$y = - \overline{c}_2^{1/2} \sigma = - 0.072(8)\, \sigma$.



\begin{references}
\bibitem{omega2}
In detail we obtain: $\omega_2 = 0.78(10)$ (conformal mapping; 
field-theoretical estimate of the FP);
$\omega_2 = 0.70(10)$ (Pad\'e-Borel; field-theoretical estimate of the FP);
$\omega_2 = 0.87(15)$ (conformal mapping; Monte Carlo estimate of the FP).
No estimate at the Monte Carlo FP could be obtained by using the 
Pad\'e-Borel method: indeed, different approximants show large fluctuations.
At the field-theoretical estimate of the 
FP $\omega_2$ has been computed in two different ways:
one can resum its perturbative expansion;
one can first resum the $\beta$-functions and then differentiate 
the resummed expressions, computing the 
stability matrix and its eigenvalues. At the Monte Carlo estimate of the
FP only the first 
method has been used.
Both methods have also been used in the analysis of $R_1$ and $R_2$ at the 
field-theoretical estimate of the FP.
\end{references}
\end{document}